%% file: main_camera_ready.tex
\documentclass[letterpaper]{article} 
\usepackage{aaai23}  
\usepackage{times}  
\usepackage{helvet}  
\usepackage{courier}  
\usepackage[hyphens]{url}  
\usepackage{graphicx} 
\usepackage{natbib}  
\usepackage{caption} 
\usepackage{algorithm}
\usepackage{algorithmic}

\usepackage{newfloat}
\usepackage{listings}
\DeclareCaptionStyle{ruled}{labelfont=normalfont,labelsep=colon,strut=off} 
\floatstyle{ruled}
\newfloat{listing}{tb}{lst}{}
\floatname{listing}{Listing}
\input{pkg}

\usepackage{url}
\title{AVCAffe: A Large Scale \underline{A}udio-\underline{V}isual Dataset of \underline{C}ognitive Load and \underline{Affe}ct for Remote Work} 

\author {
    Pritam Sarkar\textsuperscript{\rm 1, \rm 2} \quad
    Aaron Posen\textsuperscript{\rm 1} \quad
    Ali Etemad\textsuperscript{\rm 1}
}
\affiliations {
    \textsuperscript{\rm 1} Queen's University, Canada \quad
    \textsuperscript{\rm 2} Vector Institute\\
    \texttt{\{pritam.sarkar, jordan.posen, ali.etemad\}@queensu.ca}
    \\ {\textcolor{magenta}{\bf\small\texttt{https://pritamqu.github.io/AVCAffe}}}
}

\begin{document}

\maketitle

\input{AAAI/camera_ready/paper}

\subsubsection*{Acknowledgments}
We are grateful to Bank of Montreal and Mitacs for funding this research. We are thankful to SciNet HPC Consortium for helping with the computation resources. We thank Shuvendu Roy, Dept. of Electrical and Computer Engineering, at Queen's University for his collaboration during this study. We would like to further thank Prof. Kevin Munhall, Dept. of Psychology, at Queen's University for his valuable discussions at the study design stage. 

\bibliography{refs}

\clearpage
\appendix
\begin{center}
\LARGE
\textbf{Supplementary Material}
\end{center}
\input{AAAI/camera_ready/supp}

\end{document}

%% file: pkg.tex
\usepackage{longtable}
\usepackage{array}
\usepackage{booktabs}
\usepackage{color, colortbl}
\usepackage{multirow}
\usepackage{diagbox}
\definecolor{rc1}{RGB}{235,235,235}
\definecolor{rc2}{RGB}{255,255,255}
\definecolor{fade}{gray}{0.4}
\definecolor{green}{RGB}{144,238,144}

\usepackage{xcolor}
\usepackage{tikz}

\newcommand{\specialcell}[2][c]{%
\begin{tabular}[#1]{@{}c@{}}#2\end{tabular}}

\usepackage{fontawesome}

\usepackage{pifont}
\newcommand{\cmark}{\ding{51}}%
\newcommand{\xmark}{\ding{55}}%
\usepackage{graphicx}
\usepackage{amsmath}
\usepackage{booktabs}
\usepackage{enumitem}

\definecolor{blue-violet}{rgb}{0.54, 0.17, 0.89}
\definecolor{tawny}{rgb}{0.8, 0.34, 0.0}

%% file: AAAI/camera_ready/paper.tex


\begin{abstract}
We introduce AVCAffe, the first \textit{A}udio-\textit{V}isual dataset consisting of \textit{C}ognitive load and \textit{Affe}ct attributes. We record AVCAffe by simulating \textit{remote work} scenarios over a video-conferencing platform, where subjects collaborate to complete a number of cognitively engaging tasks. AVCAffe is the largest originally collected (not collected from the Internet) affective dataset in English language. We recruit $106$ participants from $18$ different countries of origin, spanning an age range of $18$ to $57$ years old, with a balanced male-female ratio. AVCAffe comprises a total of $108$ hours of video, equivalent to more than $58,000$ clips along with task-based self-reported ground truth labels for arousal, valence, and cognitive load attributes such as mental demand, temporal demand, effort, and a few others. We believe AVCAffe would be a challenging benchmark for the deep learning research community given the inherent difficulty of classifying affect and cognitive load in particular. Moreover, our dataset fills an existing timely gap by facilitating the creation of learning systems for better self-management of remote work meetings, and further study of hypotheses regarding the impact of remote work on cognitive load and affective states. 
The dataset and the supplementary material are available on the website.

\end{abstract}


\section{Introduction} \label{sec:intro}

\textit{Remote work}, also referred to as `work from home', has recently become the predominant employment paradigm for many individuals in different sectors. Dictated partially by the recent COVID-19 pandemic, and facilitated through advancements in connectivity, business communication chat platforms, and video-conferencing tools, remote work is the new reality of work for millions across the world. While this new paradigm of work has a number of advantages such as enabling social distancing and flexible hours, it brings about a number of challenges that were less common in in-person work environments. For instance, studies have shown that remote work settings could contribute to increased cognitive load and fatigues in individuals due to the following
\cite{bennett2021videoconference,fauville2021zoom,riedl2021stress}: 
(\textit{i}) back-to-back work-related meetings with minimal physical mobility in-between, (\textit{ii}) the inability to effectively perceive and transmit non-verbal expressive cues, (\textit{iii}) the need to apply intense focus on the screen with minimal variation. 

In order to better understand and manage the impact of remote work meetings on individuals, it is necessary to design and develop tools capable of quantifying factors such as cognitive load and affect in relevant settings. A key ingredient for developing such systems is the availability of related \textit{datasets} along with \textit{ground-truth information}, which could be used by machine learning and deep learning algorithms for training purposes. 
Throughout the literature, a large number of \textit{affective computing} datasets \cite{sewa,iemocap,recola,amigos,avec} have been made available, which can indeed be used to partially address the need in this area by capturing arousal, valence, and other emotion-related factors. Affective computing \cite{picard2000affective}, which is an area that aims to investigate methods and algorithms for detection \cite{sarkar2019classification,kollias2017recognition,sarkar2020self,sarkar2019self}, quantification \cite{kollias2020exploiting,kollias2021affect}, and generation of emotions \cite{kollias2018multi,kollias2020deep}, has witnessed a surge in terms of methods and performances as a direct result of the progress in deep neural networks \cite{sarkar2019classification,sarkar2020self,sarkar2019self,kollias2017recognition,kollias2020exploiting,kollias2021affect}. Nonetheless, there are currently no available datasets directed toward understanding users in \textit{remote work} settings. Moreover, there are currently no audio-visual datasets that target \textit{{cognitive load}} along side affective states. 

\begin{figure}[t]
    \centering
    \fontsize{9pt}{10pt}\selectfont
    \setlength\tabcolsep{0.35pt}
    \resizebox{1\linewidth}{!}{
    \begin{tabular}{ccccccc}

    \includegraphics[width=0.20\textwidth]{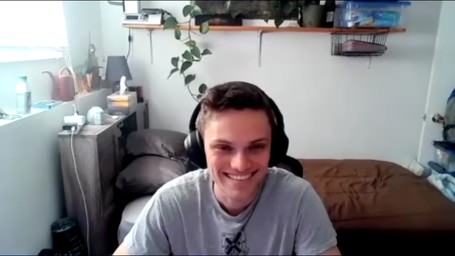} &
    \includegraphics[width=0.20\textwidth]{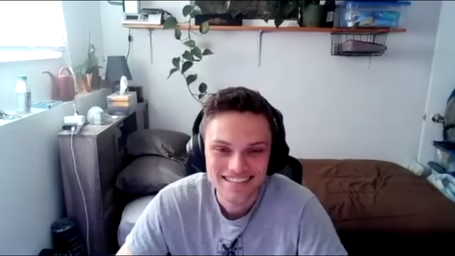} &
    \includegraphics[width=0.20\textwidth]{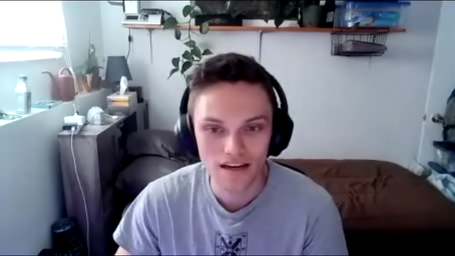} &
    \includegraphics[width=0.20\textwidth]{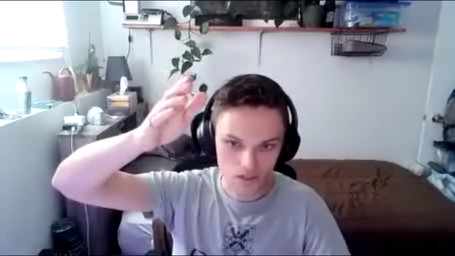} &
    \includegraphics[width=0.20\textwidth]{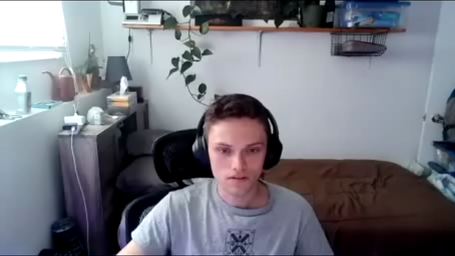} &
    \includegraphics[width=0.20\textwidth]{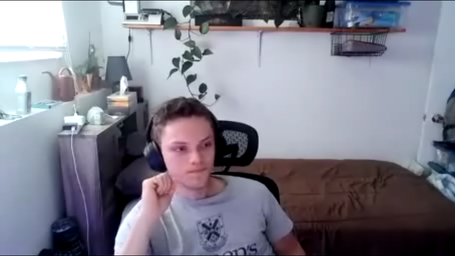} &
    \includegraphics[width=0.20\textwidth]{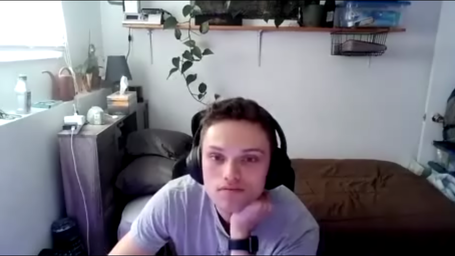} \\

    \includegraphics[width=0.20\textwidth]{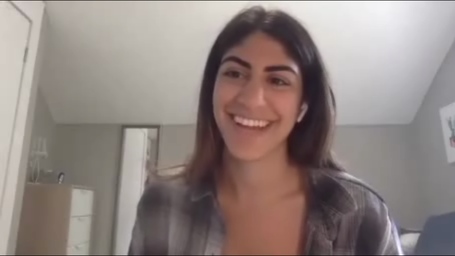} &
    \includegraphics[width=0.20\textwidth]{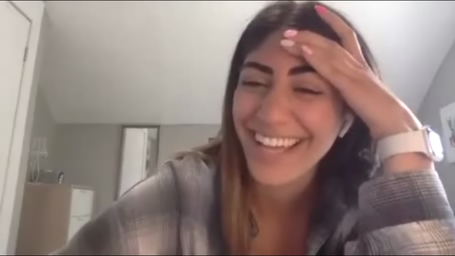} &
    \includegraphics[width=0.20\textwidth]{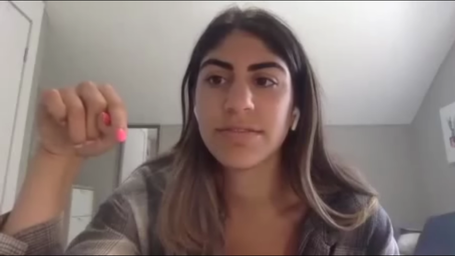} &
    \includegraphics[width=0.20\textwidth]{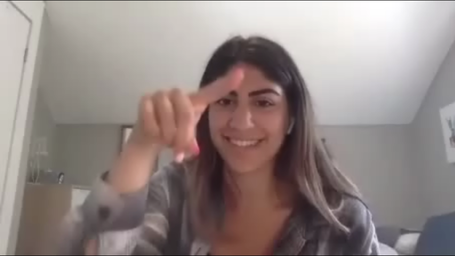} &
    \includegraphics[width=0.20\textwidth]{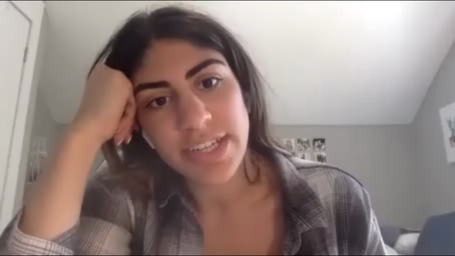} &
    \includegraphics[width=0.20\textwidth]{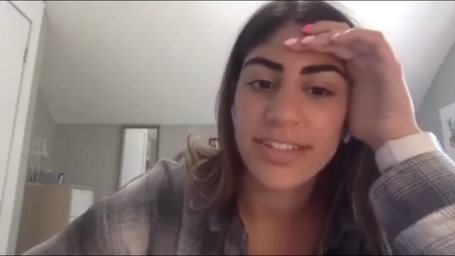} &
    \includegraphics[width=0.20\textwidth]{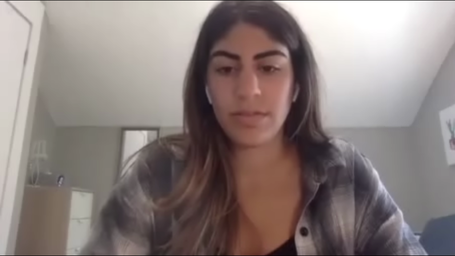} \\

    \includegraphics[width=0.20\textwidth]{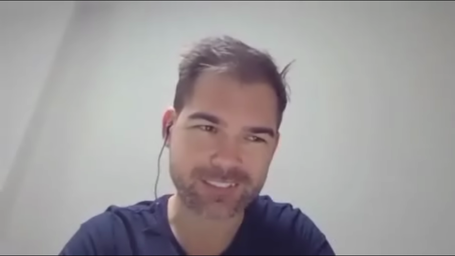} &
    \includegraphics[width=0.20\textwidth]{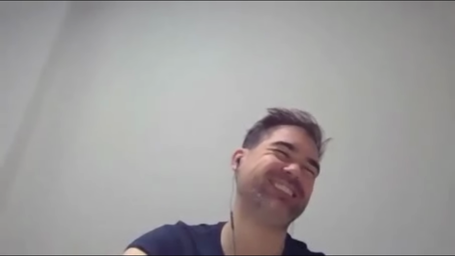} &
    \includegraphics[width=0.20\textwidth]{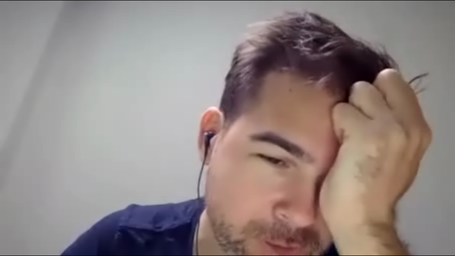} &
    \includegraphics[width=0.20\textwidth]{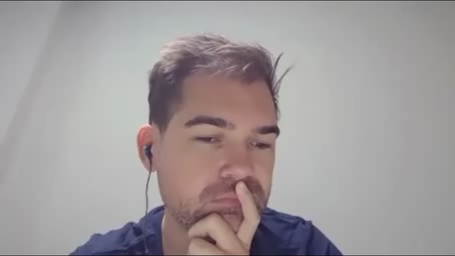} &
    \includegraphics[width=0.20\textwidth]{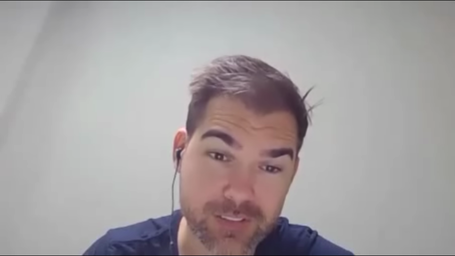} &
    \includegraphics[width=0.20\textwidth]{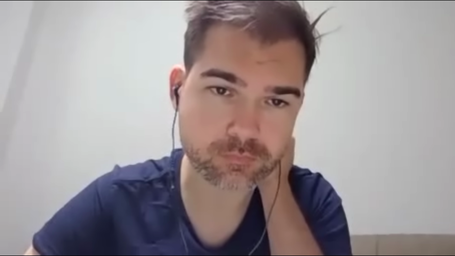} &
    \includegraphics[width=0.20\textwidth]{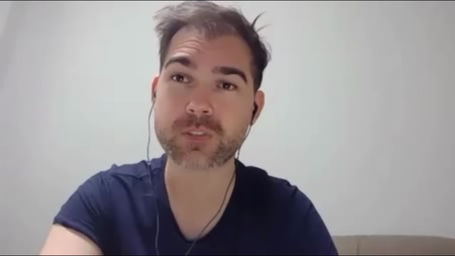} \\

    \includegraphics[width=0.20\textwidth]{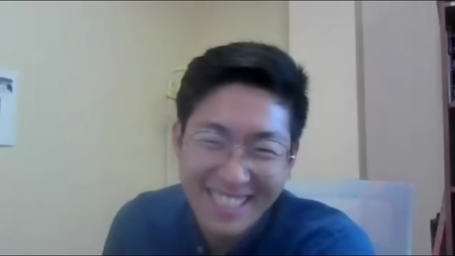} &
    \includegraphics[width=0.20\textwidth]{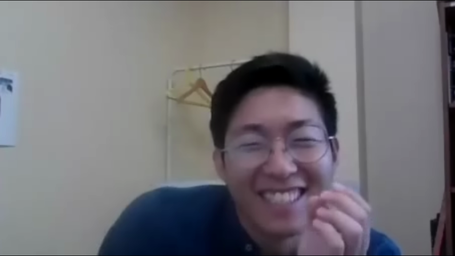} &
    \includegraphics[width=0.20\textwidth]{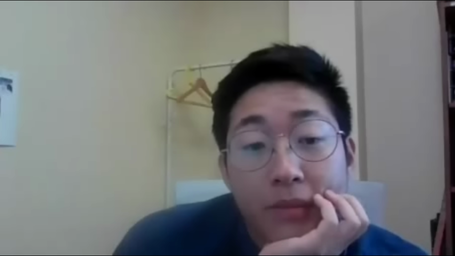} &
    \includegraphics[width=0.20\textwidth]{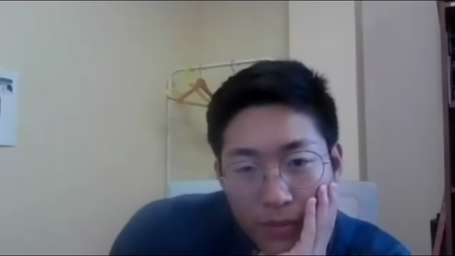} &
    \includegraphics[width=0.20\textwidth]{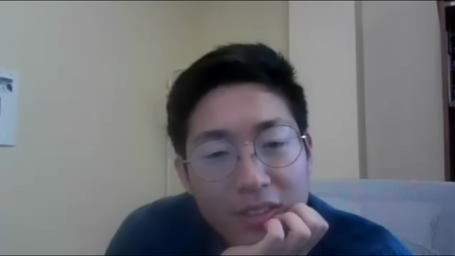} &
    \includegraphics[width=0.20\textwidth]{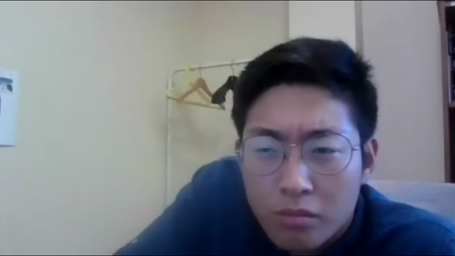} &
    \includegraphics[width=0.20\textwidth]{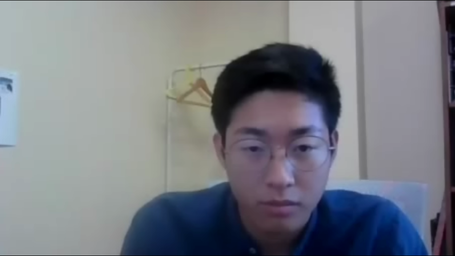} \\

    \includegraphics[width=0.20\textwidth]{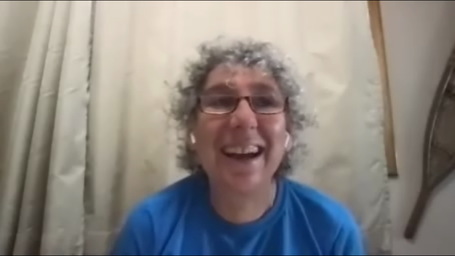} &
    \includegraphics[width=0.20\textwidth]{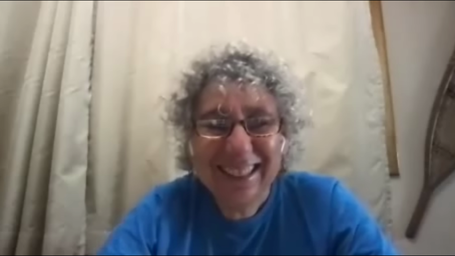} &
    \includegraphics[width=0.20\textwidth]{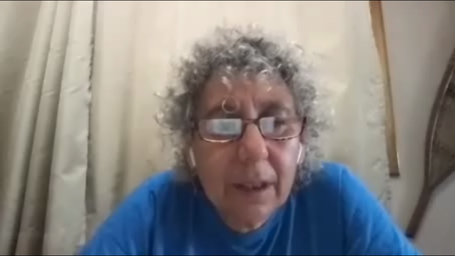} &
    \includegraphics[width=0.20\textwidth]{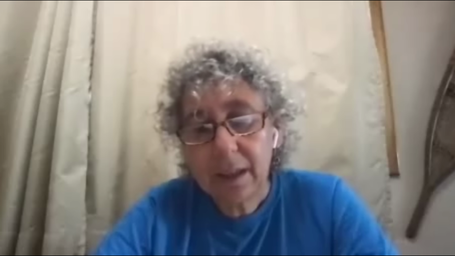} &
    \includegraphics[width=0.20\textwidth]{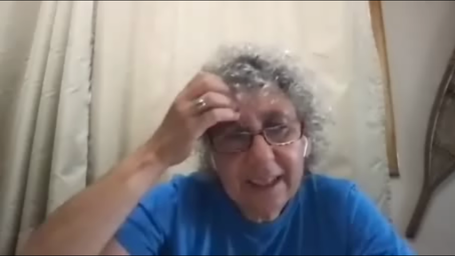} &
    \includegraphics[width=0.20\textwidth]{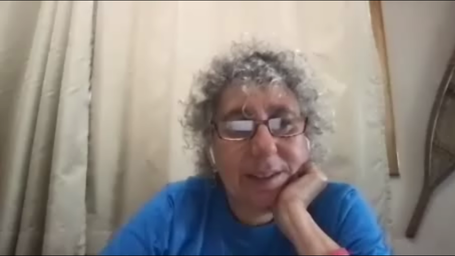} &
    \includegraphics[width=0.20\textwidth]{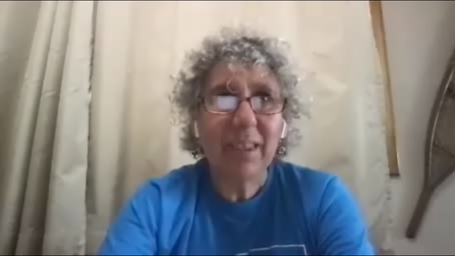} \\

    \includegraphics[width=0.20\textwidth]{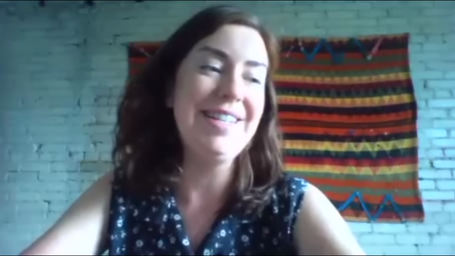} &
    \includegraphics[width=0.20\textwidth]{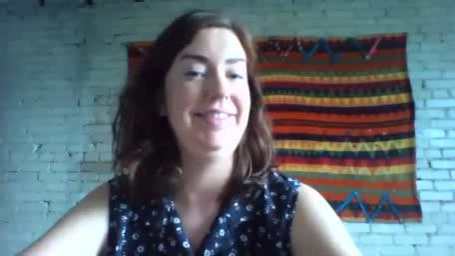} &
    \includegraphics[width=0.20\textwidth]{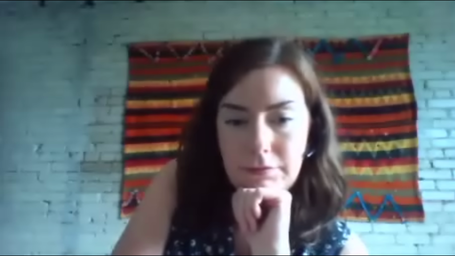} &
    \includegraphics[width=0.20\textwidth]{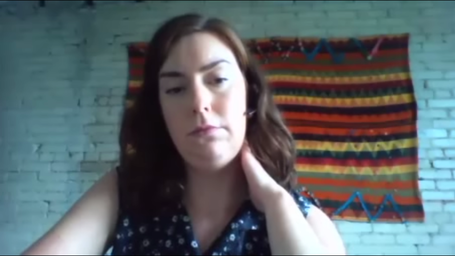} &
    \includegraphics[width=0.20\textwidth]{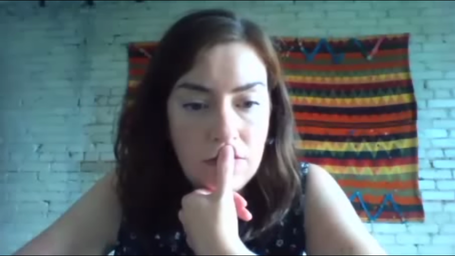} &
    \includegraphics[width=0.20\textwidth]{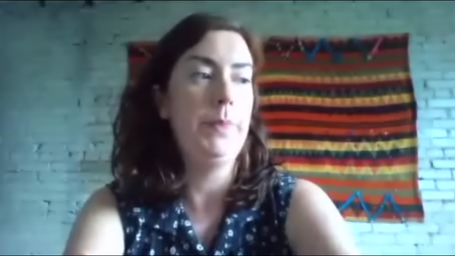} &
    \includegraphics[width=0.20\textwidth]{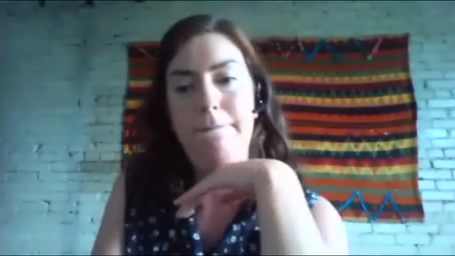} \\

    \textbf{\specialcell{Open\\Discussion}} & \textbf{\specialcell{Lighten\\the Mood}} & \textbf{\specialcell{Diapix}} & \textbf{\specialcell{Montclair\\Map}} & \textbf{\specialcell{Lost at\\the Sea}} & \textbf{\specialcell{Reading\\Comprehension}} & \textbf{\specialcell{Multi-task}} \\
    
    \end{tabular}
    }
    \caption{Sample representative frames of AVCAffe during different tasks.
    }
    \label{fig:taskwise_face}
\end{figure}

In this paper we attempt to tackle this shortcoming by developing a large-scale \textbf{A}udio-\textbf{V}isual Dataset of \textbf{C}ognitive Load and \textbf{Affe}ct (\textbf{AVCAffe}) for remote work. We procure this dataset by designing a study in which participants aim to perform a number of cognitively engaging tasks collaboratively over a video-conferencing platform. These tasks are designed, in consultations with psychologists and related literature, to elicit different cognitive and affective responses with varying intensities. In particular, our study design consists of $7$ different tasks, ranging from casual conversations, sharing jokes, finding small and precise differences between two almost similar pictures, decision making, email writing, multi-tasking and a few others. We recruit a total of $106$ participants to partake in one-on-one meetings and carry out the designated tasks together. AVCAffe consists of diverse group of participants in terms of age group ($18$ to $57$ years), ethnicity (spread over $18$ different countries of origin), profession (engineer, scientist, nurse, student, and lawyer, among several others), daily usage of online communication tools, and so on. Finally, the audio and video are recorded throughout the session, along with self-reported ground truths on a number of cognitive load and affect attributes at the end of each task.

\noindent In summary, we make the following contributions:
\begin{itemize}[noitemsep,nolistsep]
\item[$\bullet$] We create the first audio-visual dataset that includes `\textit{cognitive load}' attributes. Moreover, to the best of our knowledge, this is the first large-scale audio-visual dataset that focuses on `\textit{remote work}' settings in the context of understanding affect and cognitive load.
\item[$\bullet$] AVCAffe is the largest originally recorded audio-visual affective computing dataset (not obtained from the Internet) in English language. It consists of $106$ participants, over $108$ hours of videos, and over $58$K clips, with self-reported ground truths for affect (\textit{arousal} and \textit{valence}) and cognitive load (\textit{mental demand}, \textit{temporal demand}, \textit{effort}, \textit{performance}, \textit{physical demand}, and \textit{frustration}). 
\item[$\bullet$] We perform extensive analyses to validate our study design. We provide extensive deep learning baselines for estimating cognitive load and affect on AVCAffe, in both uni-modal and multi-modal setups.
\end{itemize}

This dataset along with supporting codes are made freely available in the 
project website to contribute to the field. The details of the dataset availability are further mentioned in Appendix \ref{supsec:availability}. 
We believe AVCAffe would be an valuable and challenging benchmark for the deep learning and affective computing research communities to accurately model cognitive load and affect, especially considering the timely context of remote work.


\section{Related Work} \label{sec:related_work}

In recent years, there has been a growing interest towards investigating human emotions and affective states across different experimental setups \cite{eNTERFACE,sewa,amigos,decaf,1438384,sarkar2019classification}, such as watching videos for mood elicitation \cite{amigos,sewa,phinnemore2021happy}, dyadic conversations \cite{msp_improv,iemocap}, and others. In this section, we summarize some of the popular affective audio-visual datasets available in the literature. Additionally, a brief overview of the existing public datasets is presented in Table \ref{tab:vid_affect_table}.

In an earlier work, an acted audio-visual dataset named IEMOCAP is created by employing $10$ skilled actors in an experimental setup of dyadic interactions~\cite{iemocap}. Similar to IEMOCAP, MSP-IMPROV~\cite{msp_improv} is an acted audio-visual database consisting of $12$ participants' audio-visual recordings, focused on understanding emotional behaviours during dyadic conversations. Other popular affective datasets include SEMAINE~\cite{semaine} and AVEC-13~\cite{avec} which are collected in human-machine interaction experiment setups. SEMAINE consists of a total of $959$ conversations between a `human' and an `operator' targeting $7$ basic emotional states, while AVEC-13~\cite{avec} consists of audio-visual recordings and self-assessed subjective depression scores from $292$ subjects.

In addition to purely affective audio-visual datasets, there are a few affective datasets such as MAHNOB-HCI~\cite{mahnob}, DECAF \cite{decaf}, and AMIGOS \cite{amigos} which are comprised of physiological signals (electrocardiogram, galvanic skin response, electroencephalogram, and others) along with the participants' audio-visual recordings. MAHNOB-HCI~\cite{mahnob} and DECAF \cite{decaf} are collected in an experimental setup where participants are asked to watch videos in order to elicit affect states. AMIGOS \cite{amigos} explores understanding of people’s emotions, personality, and mood while in groups, as well as in individual settings. In a recent work, K-EmoCon~\cite{kemocon}, recordings are acquired from a total of $32$ subjects who participated in debates on a social issue. Finally, SWEA~\cite{sewa} is an affective dataset of spontaneous behaviors where participants are asked to watch video clips in order to elicit their mental states which is followed by a discussion on the watched clips.

\begin{table}[t]
\fontsize{9pt}{10pt}\selectfont
\centering

\newcolumntype{a}{>{\columncolor{rc1}}c}
\resizebox{\linewidth}{!}{%
\begin{tabular}{lcccccp{3.1cm}}
\toprule
\multirow{2}{*}{\textbf{Database}} & \multirow{2}{*}{\textbf{\#Sub}} &  \multicolumn{3}{c}{\textbf{Annotations}} &  \multirow{2}{*}{\textbf{Size}} &   \multirow{2}{*}{\textbf{Elicitation}} \\ \cmidrule{3-5}
\rowcolor{rc2}
&& {\small \textbf{V}} & {\small \textbf{A}} & {\small \textbf{CL}} & \textbf{(hrs)} & \\
\midrule\midrule
IEMOCAP~\shortcite{iemocap} & 10 & \cmark & \cmark & \xmark & 12  & Dyadic conversation \\
MAHNOB-HCI~\shortcite{mahnob} & 27 & \cmark & \cmark & \xmark & 11  & Watching videos \\
SEMAINE~\shortcite{semaine} & 150 & \cmark & \cmark & \xmark & 80  & Human machine interaction \\
DECAF~\shortcite{decaf} & 30 &  \cmark & \cmark & \xmark & N/A & Watching movies and music videos \\
MSP-IMPROV~\shortcite{msp_improv} & 12 &  \cmark & \xmark & \xmark & 18 & Dyadic conversation \\
HUMAINE~\shortcite{humaine} & 4 &  \cmark & \cmark & \xmark & 4 & Natural and induced conversations \\
RECOLA~\shortcite{recola} & 46 &  \cmark & \cmark & \xmark & 3.5 & Online dyadic interactions \\
SEWA~\shortcite{sewa} & 398 &  \cmark & \cmark & \xmark & 44 & {Watching videos and discussion} \\
AMIGOS~\shortcite{amigos} & 40 &  \cmark & \cmark & \xmark & N/A & {Watching videos and conversation} \\
K-EmoCon~\shortcite{kemocon} & 32 & \cmark & \cmark & \xmark & 4 & Debates \\ \midrule
\textcolor{fade}{Aff-Wild~\shortcite{aff_wild}} & \textcolor{fade}{200} &  \textcolor{fade}{\cmark} & \textcolor{fade}{\cmark} & \textcolor{fade}{\xmark} & \textcolor{fade}{30} & \textcolor{fade}{Collected from YouTube} \\
\textcolor{fade}{MOSEI~\shortcite{cmu_mosi}} & \textcolor{fade}{1000} & \textcolor{fade}{\cmark} & \textcolor{fade}{\cmark} & \textcolor{fade}{\xmark} & \textcolor{fade}{66} & \textcolor{fade}{Collected from YouTube} \\ \midrule
\textbf{AVCAffe} & \textbf{106} & \cmark & \cmark & \cmark & \textbf{108} & \textbf{Remote work} \\
\bottomrule
\end{tabular}
}
\caption{A brief summary of existing public datasets are presented. 
Here, V: Valence, A: Arousal, CL: Cognitive Load.
}
\label{tab:vid_affect_table}

\end{table}

For the sake of completeness, we further briefly mention some of the popular affective datasets Aff-Wild~\cite{aff_wild}, CMU-MOSI~\cite{cmu_mosi}, Liris-Accede~\cite{liris_accede}, which are created by scraping videos from YouTube or other similar sources, unlike the datasets discussed earlier which are mostly collected in laboratory setups. Aff-Wild consists of a total of $298$ video clips of $200$ unique subjects, with a total duration of $30$ hours of data. CMU-MOSI is a collection of $23,500$ sentence utterance videos from more than $1000$ YouTube speakers, with a total duration of approx $66$ hours of data. On the other hand, Liris-Accede consists of a total of $9,800$ movie excerpts ($8$ to $12$ seconds long), with a total duration of approximately $27$ hours of audio-visual content.

\noindent\textbf{Distinctions from our work.}
We find major differences between AVCAffe and earlier works. First, none of the prior works study `cognitive load' in audio-visual modalities. To the best of our knowledge, AVCAffe is the first audio-visual dataset with cognitive load annotations. 
Specifically, we cast the problem in the context of `remote work', which has been largely under-served despite its relevance and timeliness in recent times. 
Moreover, our dataset captures data in scenarios that are much closer to in-the-wild settings than pre-planned laboratory setups. Participants frequently exhibit `spontaneous' behaviour and work `collaboratively', making AVCaffe unique amongst other datasets in the field.

Amongst all the prior works, we find RECOLA \cite{recola} to be slightly close to our approach. It attempts to investigate human emotions during online dyadic conversations, where participants perform a survival tasks in a given time period. We would like to highlight that unlike RECOLA, which consists of just one task, the experiment design of AVCAffe consists of several tasks with varied levels of difficulty to create a more realistic remote work environment. Moreover, during data collection of RECOLA, video clips are shown to participants to manipulate the states of the participants, whereas we do not perform such steps and rather aim to elicit spontaneous behaviors by the participants. Additionally, AVCAffe ($108$ hours) is a much larger dataset than RECOLA ($3.5$ hours). Lastly, we focus on investigating both affect and cognitive states, whereas RECOLA only consists of arousal and valence.


\section{AVCAffe} \label{sec:dataset}

\subsection{Study Design}
We conduct this study in an online setup, where participants and researchers join the data collection sessions through the Zoom video conferencing platform. We recruit participants for our study based on two inclusion criteria: (\textit{i}) being within the standard age range for employment, i.e., $18$ to $60$ years old; (\textit{ii}) the ability to converse in English. Additionally, we request participants not to consume alcohol, marijuana, or other substances that may severely alter their affective/cognitive states in the $12$-hour window leading to the study or during the study. To capture facial expressions or visual appearance of the participants, we recommend that participants use a computer with a webcam and a stable Internet connection. Additionally, we ask participants not to use any filters or virtual backgrounds during the sessions as it may deteriorate the video quality as well as alter the facial expressions. To collect high quality audio recordings, we recommend that participants use a headphone and avoid using any voice modulation software during the session.

Each session consists of $2$ participants, and at-least $1$ individual from the research team to moderate and facilitate the session. At the beginning of each session, we play an introductory video for the participants describing an overview and goal of the study. 
The introductory video includes short descriptions of all the tasks, a brief definition of arousal, valence, and cognitive load, as well as a few additional guidelines regarding the setup as described earlier. Next, participants are asked to fill out a pre-study questionnaire to gather some additional information about the participant pool. Questions include basic demographic information (age, sex, ethnicity), profession, and basic work-related information such as number of working hours per day, number of hours working on a computer, most frequently used mode of communication at work (video, audio, or text), and so on. 

In each session, participants complete a number of tasks. One participant is randomly assigned as `Participant A' and the other as `Participant B'. Each session typically lasts between $75$ to $90$ minutes. We record the audio-video of each participant throughout the session. The tasks are implemented using a web-based application, Qualtrics \cite{Qualtrics}. At the end of each task, we collect the self-reported ground truths to record the participants' affect and cognitive load (details are provided in Section \ref{sec:self_scores}). 

To conduct this study, we have secured ethics approval from the General Research Ethics Board at Queen’s University, Canada. The collected data are stored on secure servers, and each record is identified using an alphanumeric code. Personal information such as first and last name, email address, and others, have been discarded. Participants' consent has been collected using a secure web-based form. We seek $2$ types of approval from the participants: (\textit{i}) approval to participate in the study, which should be `Yes' in order for the study to proceed; (\textit{ii}) approval to use the participants' image/video/audio in articles, publications, or accompanied media content, where participants could opt for `No', but still participate in our study. Note that participants' photos used in this manuscript are only taken from those participants who have explicitly approved the use of their images in publications.

\subsection{Task Design} \label{sec:tasks}
Our goal is to design a study protocol that closely resemble remote work and meetings. To facilitate this concept, we devise a series of tasks with varied levels of difficulty, eliciting cognitive load and affect in a controlled experimental setup. Our study design requires $2$ participants to collaborate and communicate over a video conferencing platform to successfully complete the tasks. Each session is composed of $7$ tasks, which are designed for a total duration of approximately $1$ hour. A brief description of each task is presented below. Further discussions on the rationale behind choosing each task is discussed later in Section \ref{sec:design_consideration}.

\noindent\textbf{A. Open discussion.}
In this task, participants discuss a topic of their choice. We provide a set of non-personal potential topics along with a few suggested discussion points for each topic. However, other topics are allowed to be chosen upon agreement between both the participants. The suggested topics include, (\textit{i}) movies and TV series, (\textit{ii}) games or sports, (\textit{iii}) books, (\textit{iv}) work, (\textit{v}) hobbies, passions, or creative interests. A duration of $7.5$ minutes is allocated for this task. 

\noindent\textbf{B. Lighten the mood.}
In this stage, participants are asked to share some interesting or humorous incidents, or alternatively tell jokes to the other participant. We provide a few pre-written jokes to each participant in case they can't think of any. Similar to \textit{open discussion}, $7.5$ minutes is allocated for this task.

\noindent\textbf{C. Diapix.}
In this task participants are given two highly similar pictures \cite{diapix}, in which they need to find and mark the differences. Each participant is only given one of the two pictures and they have to work together to find the differences between the pair of pictures through verbal communication. Participants are asked to find a total of $10$ differences and mark them in under $10$ minutes.

\noindent\textbf{D. Montclair map.}
During this task, participants complete a map-matching task \cite{montclair} with one another. The two participants receive nearly identical maps that have paths drawn on them from a starting point around/through various landmarks leading to a finish point. Without seeing each other’s maps, the participants need to communicate with each other to locate the missing landmarks on both maps. This task needs to be completed in under $2.5$ minutes, and $2$ of such tasks need to solved consecutively. 

\noindent\textbf{E. Lost at sea.}
In this stage, participants are presented with a hypothetical situation \cite{lost_at_sea} where participants are on a chartered yacht with friends, crew, and a captain, for a holiday trip. Due to unfortunate circumstances, a fierce fire breaks out on the ship and all the crew members and the captain are lost. The task of the participants is to prioritize $15$ items which need to be saved for their survival in the middle of the sea. The items need to be selected unanimously and ranked under $10$ minutes. 

\noindent\textbf{F. Reading comprehension.}
In this task, one participant (active) reads a given a short passage from \cite{reading_comp}, followed by answering a few related questions asked by the other participant (passive). The roles of the $2$ participants are then reversed. The passages and questions are both provided by us. $5$ minutes is allocated to each round, where $2$ minutes are to be used for reading and $3$ minutes are allocated for answering.

\noindent\textbf{G. Multi-task.}
The last component of the session is a multi-task scenario, as often encountered in our day-to-day work. In this task, both participants write an email based on a few given points. They are both tasked with interrupting each other by asking a few questions unrelated to the email. The questions are provided to the participants at regular intervals using popup messages. The interrupting questions are simple arithmetic problems (e.g.,, summation, multiplication, or division of $2$ numbers), as well as synonyms/antonyms questions related to common English words.

\subsection{Ground Truths} \label{sec:self_scores}

\begin{figure}[!t]
    \centering
    \fontsize{9pt}{10pt}\selectfont
    \setlength\tabcolsep{0.5pt}
    \resizebox{0.45\textwidth}{!}{%
    \begin{tabular}{lrrr}
    & \textbf{Open discussion} & \textbf{Montclair map} & \textbf{Multi-task} \\
    \begin{tikzpicture}
    \draw [white] (0,0) rectangle node{\large \textcolor{black}{\textbf{Participant A}}} (2.4cm,2.7cm);
    \end{tikzpicture}    & \includegraphics[width=0.16\textwidth]{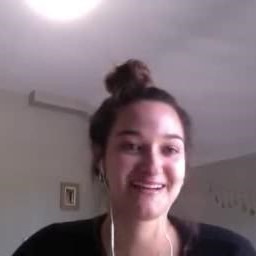} 
    & \includegraphics[width=0.16\textwidth]{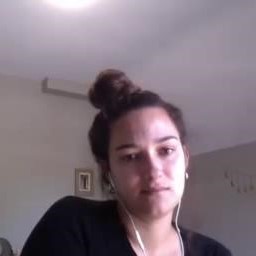} 
    & \includegraphics[width=0.16\textwidth]{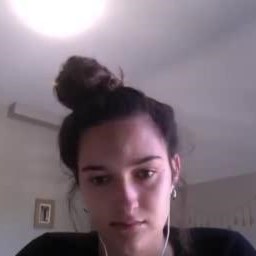} 
     \\
    \textbf{Arousal}& Wide-awake & Excited & Wide-awake\\
    \textbf{Valence}& Pleasant & Unsatisfied & Pleased \\
    \textbf{Effort}& 8 & 17 & 16 \\
    \textbf{Mental demand}& 3 & 12 & 13 \\
    \textbf{Temporal demand}& 0 & 20 & 11 \\
    \begin{tikzpicture}
    \draw [white] (0,0) rectangle node{\large \textcolor{black}{\textbf{Participant B}}} (2.4cm,2.7cm);
    \end{tikzpicture}
    & \includegraphics[width=0.16\textwidth]{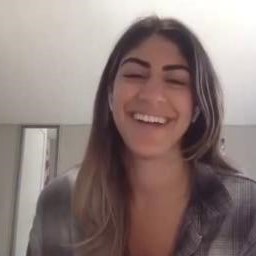} 
    & \includegraphics[width=0.16\textwidth]{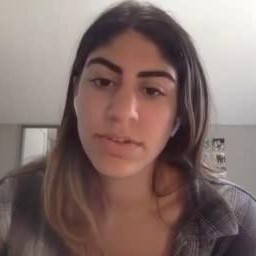} 
    & \includegraphics[width=0.16\textwidth]{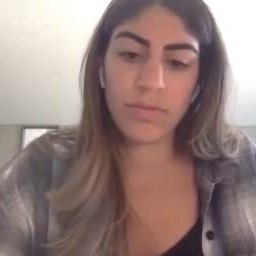} 
     \\
    \textbf{Arousal}& Wide-awake & Excited & Wide-awake\\
    \textbf{Valence}& Pleasant & Pleased & Pleasant \\
    \textbf{Effort}& 2 & 16 & 12 \\
    \textbf{Mental demand}& 4 & 12 & 15 \\
    \textbf{Temporal demand}& 3 & 16 & 6 \\

    \end{tabular}
    }
    \caption{Sample clips along with self-reported affect and cognitive load scores during different task.
    }
    \label{fig:session}
\end{figure}
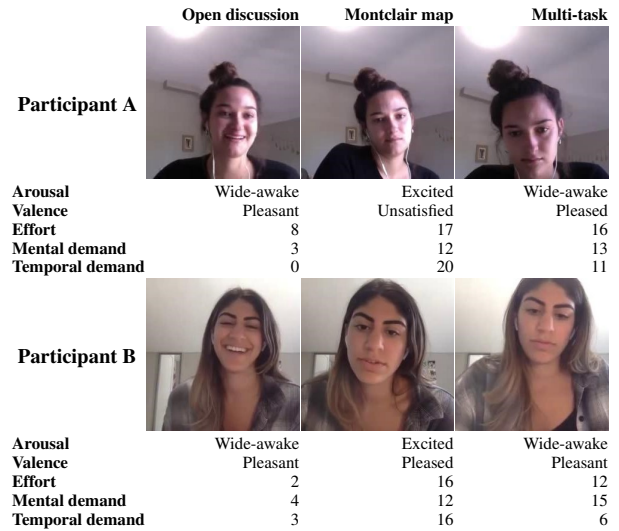

We use the NASA Task Load Index (NASA-TLX) \cite{nasa_tlx} and Self Assessment Manikin (SAM) \cite{sam} scales to collect cognitive load and affect scores, respectively. Following the protocols laid out in \cite{nasa_tlx}, cognitive load scores are collected on a scale of $0$-$21$, across mental demand, physical demand, temporal demand, performance, effort, and frustration categories. Additionally, participants report their arousal and valence scores \cite{sam} on a scale of $0$-$4$. For the sake of completeness, we present the questionnaire used to collect the responses in 
the Appendix \ref{supsec:questionnaire}.
We present sample ground truths reported by two of the participants in Figure \ref{fig:session}.

We note that it has become common practice \cite{cmu_mosi,sewa,liris_accede} to annotate arousal and valence through `external annotators' either via crowdsourcing or experts, especially when there is no access to self-reported ground truths, e.g., when videos are scraped from the Internet or movie excerpts. However, we find two issues with this approach. First, cognitive load attributes such as mental demand, effort, and temporal demand, are highly complex mental states, and we could not come across any literature to support that cognitive load can be annotated through external annotators, at least to a reasonable degree of confidence. Second, it has been pointed out \cite{liris_accede,zeng2018facial} that this approach suffers from poor intra-annotator (ratings from the same annotator) as well as inter-annotator (ratings from different annotators) inconsistency due to the subjective nature of \textit{emotion}. Therefore, we intentionally collect and rely on self-reported scores for both the affect and cognitive load ground truth values, and do not employ external annotators to annotate any part of the data.

\subsection{Design Considerations} \label{sec:design_consideration}
We include the \textit{open discussion} and \textit{lighten the mood} activities to induce positive emotions and enable participants to become comfortable with each other, which in turn could enable more effective collaborations. We carefully choose different tasks to target varying levels of cognitive and affective states at different stages of the experiment. For instance the \textit{Montclair map} task is specifically aimed at higher temporal demand given the short amount of allotted time, while the \textit{reading comprehension} and \textit{multi-task} activities are included to induce higher mental demand. Our analysis provided in Section~\ref{sec:analysis} demonstrates that various types and levels of cognitive and affective states have indeed been induced at different stages of the experiment. It should be noted that the tasks appear in the same order as mentioned in Section \ref{sec:tasks} for all the participants to induce incremental cognitive load in a controlled manner. We therefore do not shuffle the order of the tasks for different participant pairs. 

As mentioned earlier, we collect self-reported ground truths from the participants at the end of each task. We note that this setup results in sparse annotations for the corresponding audio-visual recordings. A potential alternative would have been to collect ground truth values at shorter interval, in between tasks. However, we noticed in our dry runs that more frequent interruptions would be too disruptive to the flow of tasks and distract the participants from the actual problems designed to be solved.

\subsection{Data Pre-processing } \label{sec:Data Pre-processing}

First, using the raw videos collected at the end of each session, we prepare a separate video recording of each participant for each task. These \textit{long videos} are typically between $2.5$ minutes to $10$ minutes in length, based on the duration of the tasks. To locate affect and cognitive load attributes at a more temporally granular scale, we further segment the \textit{long videos} into \textit{short video segments} with average duration of $6$ seconds based on the speaker's utterances, using the silence detection algorithm \cite{Pydub}.
Finally, we resize the frame resolution to their shorter side at $256$ pixels. 
We set the video sampling rate at $25$ FPS and the audio sampling rate at $44.1$ kHz. 

\subsection{Data Split}
We separate the samples into train and validation/test splits maintaining even distributions of age, gender, and ethnicity. We set aside $20$ participants for validation and rest of the $86$ participants are used for training. Moreover, to avoid any information leakage between the train and validation splits, participants from the same session always reside in the same split, (either in train or validation).

\begin{table}[]
    \centering
    \fontsize{9pt}{10pt}\selectfont
    \resizebox{0.45\textwidth}{!}{%
    \begin{tabular}{p{3cm}p{3cm}p{3.5cm}}
    \toprule
    \textbf{\# Subjects.} $106$
    & \textbf{Duration.} $108$~hrs.
    & \textbf{\# Clips.} $58,112$ \\ \midrule
     \multicolumn{2}{p{6cm}}{
    \textbf{Video.} Resolution:~$456\!\times\!256$;
                    Rate: $25$fps;
    }
    & \textbf{Audio.} Freq.: $44.1$KHz.
    \\ \midrule
    \multicolumn{3}{p{7.6cm}}{
    \textbf{Gender.} Male:~$52$, Female:~$53$, Non-Binary:~$1$.} \\ \midrule
    \multicolumn{3}{p{10cm}}{
    \textbf{Age:}
    $18$ to $20$~:~$8$; 
    $21$ to $30$~:~$75$;
    $31$ to $40$~:~$17$;
    $41$ to $50$~:~$2$;
    $51$ to $60$~:~$4$. 
    } \\ \midrule
    \multicolumn{3}{p{10.5cm}}{
    \textbf{Countries of origin:}
    Bangladesh(1), Brazil(2), Canada(67), China(3), Ecuador(1), Egypt(1), Germany(1),
    Hong Kong(1), India(11), Iran(4), Ireland(1), Jordan(1), Mexico(4), Nigeria(2),
    Pakistan(2), Sweden(1), USA(2), Vietnam(1)
    } \\ \midrule
    \multicolumn{3}{p{10.5cm}}{
    \textbf{Ground truths.}
    Arousal, Valence, Mental Demand, Temporal Demand, Effort, Physical Demand, Performance, and Frustration
    } \\ \bottomrule
    \end{tabular}
    }
    \caption{Key statistics of AVCAffe.}
    \label{tab:dataset_stats}

    \vspace{10pt}
    \includegraphics[width=\linewidth]{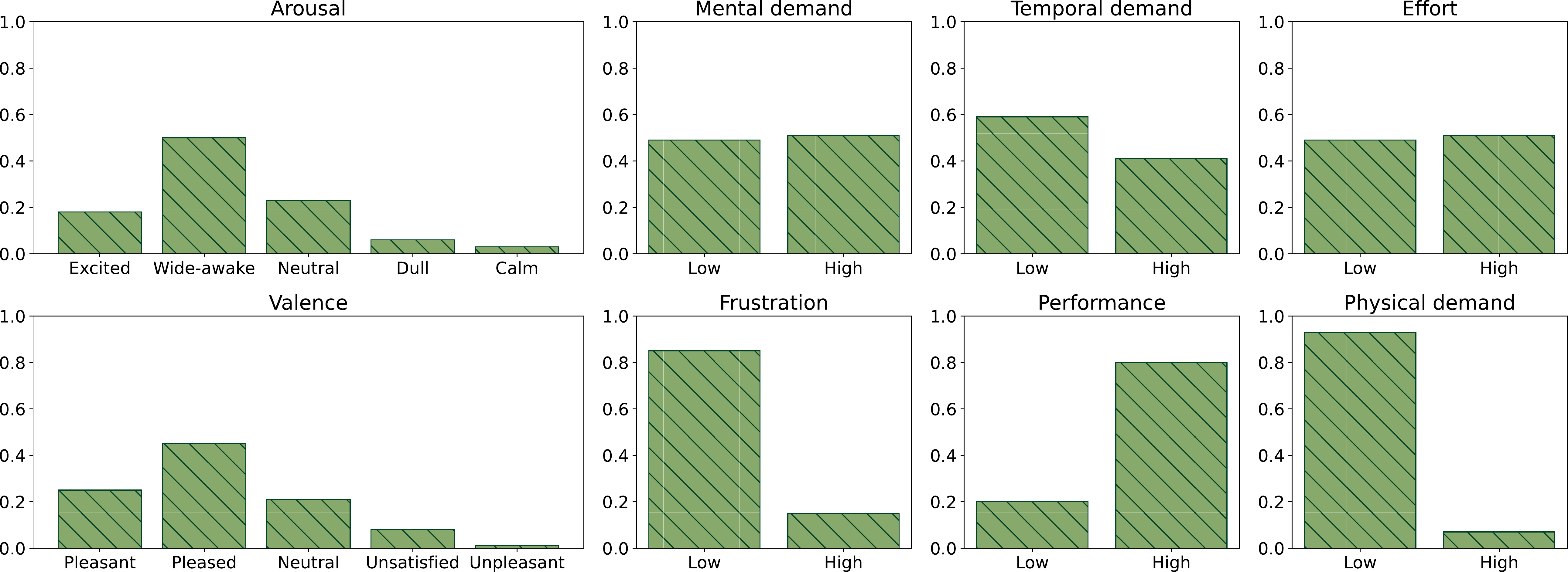}\\
    \caption{Class distributions of AVCAffe.}
    \label{fig:class_dist}
\end{table}

\subsection{Statistics}
In this study, we recruit a total of $106$ participants, spread over $18$ different countries. Our participant pool consists of $49\%$ male, $50\%$ female, and $1\%$ non-binary participants. Out of the total participants, around $60\%$ are from North America, and the rest of the population belongs to India, Iran, and several other countries. Additionally, the data population is also spread over a wide range of age groups, specifically between $18$-$57$ years. Lastly, AVCAffe consists of approximately $58$K video clips, equivalent to $108$ hours of video recordings and their corresponding affect and cognitive load labels. The key statistics are highlighted in Table \ref{tab:dataset_stats}. 

\begin{figure*}[!t]
    \centering
    \begin{tabular}{cc}
    \includegraphics[width=0.47\textwidth]{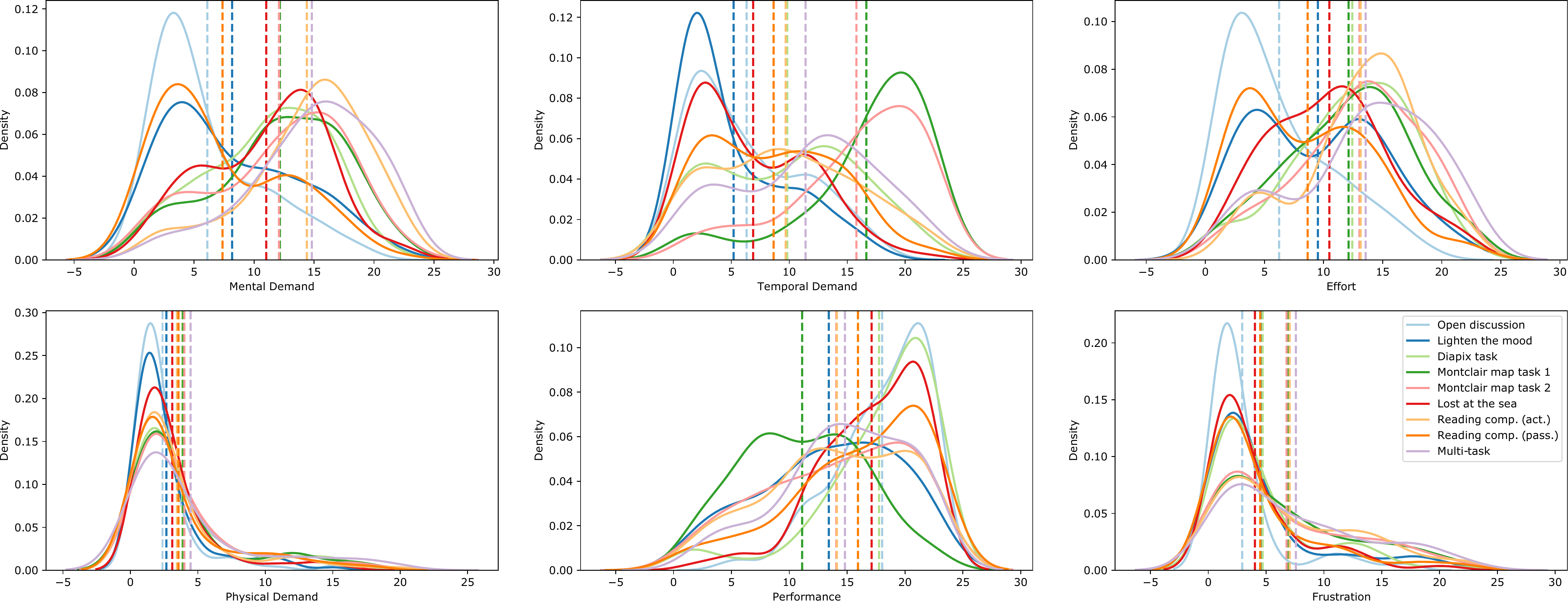} 
    &
    \includegraphics[width=0.51\textwidth]{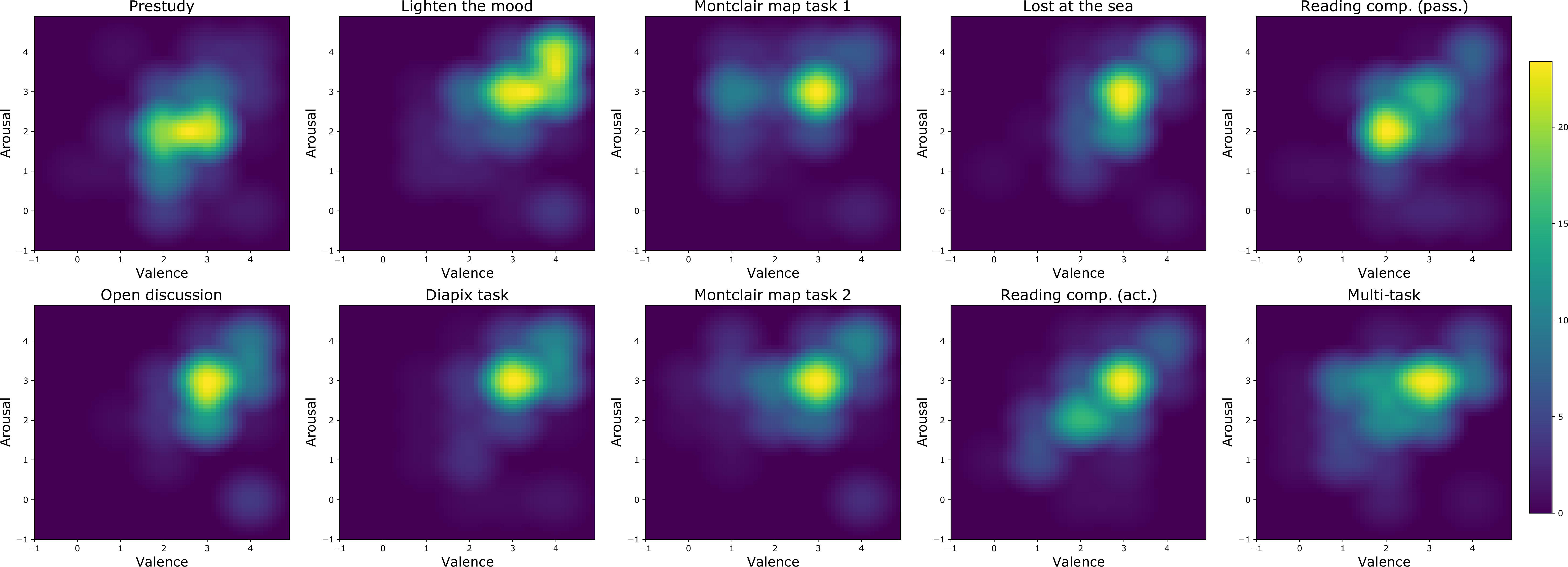} 
    \end{tabular}
    \caption{Left:
    We present the density plots of self-reported cognitive load scores, each color refers to an individual task.
    Right: We present the affect scores projected in a $3$-d plane, where
    the colors denote population density, yellow being the most dense.
    }
    \label{fig:dist_cog_aff}

\end{figure*}

We present the class distributions of all the ground truth labels in Figure \ref{fig:class_dist}. We notice that `wide-awake' and `pleased' which are the second-highest choices of arousal and valence respectively, are the dominating selections.
We observe more or less balanced distributions for mental demand, temporal demand, and effort. However, in case of frustration and physical demand, the majority ($\geq 80\%$) of the videos are categorized as `low', which is expected as our tasks are neither meant for high physical demand nor meant to elicit high levels of frustration. Similarly, we find `high' as the dominating class for performance, which is because participants are mostly successful in finishing these tasks.

\section{Analysis and Evaluation} \label{sec:analysis_eval}

\subsection{Analysis} \label{sec:analysis}
We perform an in-depth analysis of self-reported ground truths and present the results in Figures \ref{fig:dist_cog_aff} and \ref{fig:score_corr}. Our analysis helps us validate the success of our study design, indicating that different tasks are able to induce varying degrees of cognitive and affective states amongst the participants throughout the session. Our detailed findings are as follows.

\subsubsection*{Cognitive Load.}
While analyzing cognitive states, we present the density plots of each cognitive load attribute such as, mental demand, effort, and temporal demand in Figure \ref{fig:dist_cog_aff}. We find distinct shifts in cognitive load 
over time, specifically for \textit{mental demand}, \textit{effort}, and \textit{temporal demand}. We notice that the participants' \textit{mental demand} is fairly low during the \textit{open discussion} and \textit{lighten the mood}. On the other hand, participants show higher mental demand during tasks like \textit{multi-task} and \textit{reading comprehension (active)}. Moreover, participants show moderate mental demand during other tasks like \textit{diapix}, \textit{montclair map}, and \textit{lost at the sea}. It should be noted that, these outcomes are completely in line with our intended design considerations.
Next, while analysing temporal demand, we find that participants experience higher temporal demand during \textit{montclair map}; moderate temporal demand during \textit{multi-task} and \textit{reading comprehension}; and low temporal demand during tasks like \textit{open discussion} and \textit{lighten the mood}. 
Participants report low effort required to complete tasks like \textit{open discussion} and \textit{reading comprehension (passive)}, whereas high effort scores are reported for \textit{multi-task} and \textit{reading comprehension (active)}.
In case of other cognitive load attributes such as \textit{physical demand} and \textit{frustration}, we do not notice any considerable shifts across different tasks, and participants always report low score on these attributes. 
These findings coincide with our intended design as our study is not meant to evoke \textit{frustration}. Moreover, due to the fact that all of these tasks are computer-based, minimal \textit{physical demand} is required.
Lastly, in case of \textit{performance}, we notice minimal shift in self-reported scores across different tasks. On average, participants report moderate to high in terms of successful completion of the tasks, i.e., \textit{performance}.

\subsubsection*{Affect.}
To analyse the self-reported affect scores, we project arousal and valence responses onto a 3D valence-arousal space, presented in Figure \ref{fig:dist_cog_aff}. We notice considerable shifts in self-reported responses across different stages of the experiment. For example, at the beginning of the sessions, i.e., the \textit{Prestudy}, the majority of the participants report `neutral' in terms of both arousal and valence. Next, in case of \textit{Open discussion} and \textit{Lighten the mood}, a clear shift in arousal and valence is noticed from the center to the first quadrant of the valence-arousal space. Specifically during \textit{Lighten the mood}, participants report high arousal (`excited') and high valence (`pleasant'). Interestingly, during the next task (\textit{Diapix}), we notice a downward shift for both arousal and valence, i.e., `excited' to `wide-awake' for arousal, and `pleasant' to `pleased' for valence. Moreover, during \textit{Montclair map}, the majority of the responses remain the same as the previous task. However, we notice some participants experience `neutral' in the case of arousal, and `unsatisfied' in the case of valence. Similarly, during \textit{reading comprehension (active)} and \textit{multi-task}, we observe further shifts in affect response towards the third (`dull' and `unsatisfied') and second (`wide-awake' and `unsatisfied') quadrants of the valence-arousal space. Such prominent shifts in valence-arousal space during different tasks indicate that our study design successfully targets different affective states at different stages of the session. 

\subsubsection*{Inter-label relationships.} 
\begin{figure}[t]
    \centering
    \includegraphics[width=\linewidth]{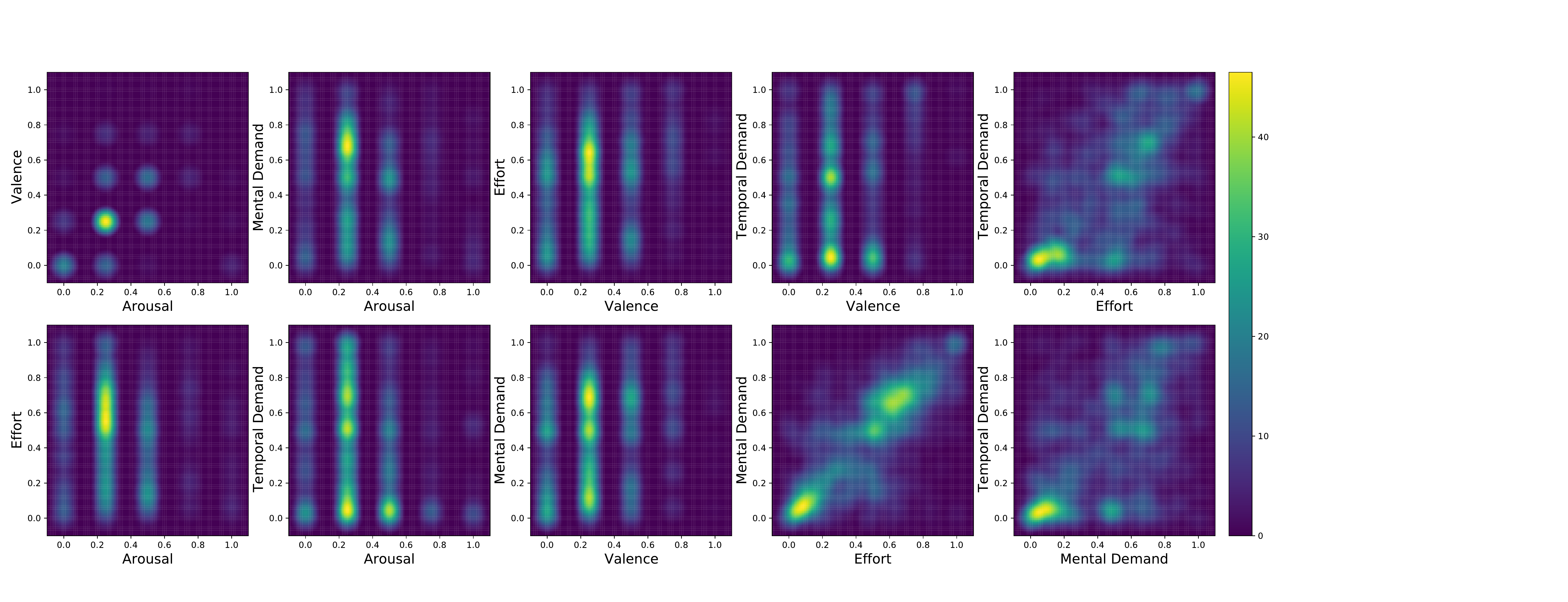}
    \caption{Qualitative analysis on inter-label relationships.}
    \label{fig:score_corr}
\end{figure}

Lastly, we aim to explore the relationships between the prominent affect and cognitive load attributes, namely arousal, valence, mental demand, effort, and temporal demand.
To investigate this, we project the normalized self-reported ground-truths onto a 3D space as shown in Figure \ref{fig:score_corr}. We observe a strong positive correlation between effort and mental demand, indicating that with increasing amounts of effort, participants experience higher mental demand, and vice-versa. Additionally, some degree of positive correlation is noticed between temporal demand and effort, as well as between temporal demand and mental demand. Interestingly, we do not observe strong correlations between cognitive load and affect attributes. This indicates that our dataset has been able to successfully capture unique information beyond the arousal and valence classes. We further provide a quantitative analysis in the Appendix \ref{suppsec:inter-label}.

\begin{table*}[t]
        \fontsize{9pt}{10pt}\selectfont
        \centering
        
        \resizebox{0.85\linewidth}{!}{%
        \begin{tabular}{ccccccccccccc}
        \toprule
        \multirow{2}{*}{\textbf{Audio}} & \multirow{2}{*}{\textbf{Visual}}
         & \multirow{2}{*}{\textbf{\#Params}}
        & \multicolumn{2}{c}{\textbf{Mental D.}} & \multicolumn{2}{c}{\textbf{Effort}} & \multicolumn{2}{c}{\textbf{Temporal D.}} &
        \multicolumn{2}{c}{\textbf{Arousal}} & \multicolumn{2}{c}{\textbf{Valence}}\\ 
        &&& \textbf{Short} & \textbf{Long} & \textbf{Short} & \textbf{Long} & \textbf{Short} & \textbf{Long} & \textbf{Short} & \textbf{Long} & \textbf{Short} & \textbf{Long}  \\ \midrule \midrule
        \multicolumn{2}{c}{{Random Classifier}} & - & 51.3&45.6 & 48.0&43.6 & 35.1&33.9 & 32.9&26.2 &  34.0&30.3  \\ \midrule
        VGG-16 & \xmark & 14.7M & \underline{58.8}&\underline{61.2} & \underline{58.8}&\underline{62.1} & 57.9&\underline{56.7} & \underline{38.3}&\underline{36.1} &  \underline{40.3}&\underline{39.1} \\ 
        ResNet-18 & \xmark & 11.2M & 58.2&60.7 & 57.0&60.8 & \underline{58.2}&54.4 & 38.1&30.4 &  39.3&36.3 \\ \midrule 
        
        \xmark & MC3-18 & 11.7M & 60.4&\underline{61.0} & 61.4&63.8 & \underline{60.0}&\underline{59.4} & \underline{41.4}&\underline{34.0} &  \underline{42.0}&38.8 \\ 
        \xmark & ResNet3D-18 & 33.4M & 59.3&59.0 & 61.0&62.7 & 58.5&57.9 & 37.8&30.9 &  41.9&\underline{39.5} \\ 
        \xmark & R(2+1)D-18 & 31.5M & \underline{60.5}&59.6 & \underline{\textbf{65.5}}&\underline{67.7} & 59.6&54.9 & 39.7&33.3 &  38.7&34.9 \\ \midrule 
        
        VGG-16 & MC3-18  & 47.4M & 59.4&60.2 & 59.7&66.2 & 60.8&61.4 & 41.3&38.9 &  40.3&\underline{\textbf{41.7}} \\ 
        VGG-16 & ResNet3D-18   & 69.1M & \underline{\textbf{65.0}}&\underline{\textbf{65.6}} & 59.7&60.5 & 59.2&60.3 & 40.7&37.3 &  \underline{\textbf{43.9}}&39.4 \\ 
        VGG-16 & R(2+1)D-18 & 67.2M & 60.1&64.7 & 59.7&\underline{\textbf{69.4}} & 60.4&\underline{\textbf{66.7}} & 42.1&\underline{\textbf{40.5}} &  41.1&39.5 \\ 
        ResNet-18 & MC3-18  & 43.9M & 61.3&60.2 & 59.4&62.1 & 58.8&57.7 & 42.4&36.0 &  41.4&39.2 \\ 
        ResNet-18 & ResNet3D-18   & 65.6M & 58.8&61.2 & 60.7&64.4 & \underline{\textbf{61.2}}&61.7 & 42.6&35.1 &  39.8&39.1 \\ 
        ResNet-18 & R(2+1D)-18 & 63.7M & 60.4&62.7 & \underline{60.8}&61.1 & 58.6&59.0 & \underline{\textbf{44.0}}&39.5 &  40.9&37.7 \\ 
        \bottomrule
        \end{tabular}
        }
        \caption{Baselines on AVCAffe are presented. 
        The best F1-scores in each subcategory (audio-only/visual-only/audio-visual) are underlined and best scores of each label are highlighted in bold. Here, Random Classifier refers to a randomly initialized classifier with no training which serves as a reference point to understand the performance of different models. 
        }
        \label{tab:results}
\end{table*}


\subsection{Benchmark Evaluation} 

\subsubsection{Training.}
We present exhaustive deep-learning baselines on AVCAffe in both uni-modal and multi-modal setups. We use a wide variety of architectures as video and audio backbones. In particular, we use R(2+1)D \cite{r2plus1d}, ResNet3D \cite{3dresnet}, and MC3 \cite{r2plus1d} as the video backbones. Additionally, ResNet \cite{resnet} and VGG \cite{vgg} are used as the audio backbones. In case of uni-modal training, we use the embeddings obtained from the backbones and pass them through a fully-connected layer. In case of multi-modal training, we perform late fusion \cite{zhang2016multimodal} by concatenating the embeddings obtained from audio and video backbones, followed by passing them through an MLP head. 
As mentioned earlier, the self-reported arousal and valence scores are originally collected on $5$-point scales which correspond to $5$ distinct classes. Therefore, we formulate this problem as a multi-class classification task. The self-reported scores for cognitive load are originally collected on a scale of $0$-$21$. We empirically find it quite challenging to accurately model such level of granularity for cognitive load due to it's inherent complexity. Hence, to make the task simpler, we formulate this problem as binary classification. Following the NASA-TLX scale \cite{nasa_tlx}, the self-reported cognitive scores greater than $10$ are marked as {High} and less than or equals to $10$ are marked as {Low}. Moreover, to create the baselines, we do not use \textit{frustration}, \textit{physical demand}, and \textit{performance}, as these three attributes do not show enough variance in our collected dataset (see our discussion in Section \ref{sec:analysis}). Next, to train the framework, we feed $2$ seconds of audio visual segments to the networks. In particular, for audio segments we generate mel-spectrograms and for visual segments we obtain the face crops before feeding to the network. We use cross-entropy loss to train the network, specifically binary cross-entropy loss for cognitive load and categorical cross-entropy for affect attributes. Additional details on training are provided in the Appendix \ref{supsec:baseline}.

\subsubsection{Evaluation Protocol.}
We evaluate AVCAffe on a total of $5$ affect (arousal and valence) and cognitive load (mental demand, effort, and temporal demand) attributes. We perform multi-class and binary classification for affect and cognitive load respectively. We evaluate the models at two levels, (\textit{i}) \textit{short video segments} (duration of $6$ seconds) and (\textit{ii}) \textit{long videos} (duration of $2$ to $10$ minutes). To evaluate on \textit{short video segments}, we uniformly sample $3$ clips per sample and report the F1-score by averaging the clip level predictions. Next, to obtain the prediction correspond to \textit{long videos} we average the predictions correspond to the \textit{short video segments}. 
Following a standard practice \cite{iemocap,sewa,aff_wild}, we use the weighted F1-measures as the evaluation metric because of it's robustness towards imbalanced class distribution. Please note, we do not apply any augmentations during validation.

\subsubsection{Results.}
The performances of the baselines are presented in Table \ref{tab:results}. We notice that in overall multi-modal networks outperform the uni-modal variants in both \textit{short video segments} and \textit{long videos} on all the attributes except \textit{Effort} on \textit{short video segments}. In particular, we notice that visual features work considerably well in predicting \textit{Effort} on \textit{short video segments} and even outperform the multi-modal variants.
Additionally, while comparing between audio-only vs. visual-only, we find that in most of the cases visual-only models perform better. We empirically find that using the face-crops instead of the whole frames is helpful to extract better representations from the visual streams. Additionally, we notice that our multi-modal baselines perform relatively better in classifying \textit{affect} attributes on the \textit{short video segments} in comparison to the \textit{long videos}. For example, the best multi-modal scores achieved on arousal and valence are $44.0$ and $43.9$ on \textit{short video segments} and $40.5$ and $41.7$ on \textit{long videos}.
However, a different trend is noticed on the \textit{cognitive load} attributes. For example, our best model on \textit{temporal demand} achieves F1-scores of $66.7$ on \textit{long videos} and $61.2$ on \textit{short video segments}. Similar trend is also noticed in the case of Effort. This interesting finding may open the door to future lines of inquiry into the differences between affect and cognitive load both from a human perception and from an affective computing perspective. 


\section{Summary} \label{sec:discussion}

We present a novel audio-visual database of cognitive load and affect collected in a setup resembling `remote work'. To the best of our knowledge AVCAffe is the first audio-visual dataset comprised of \textit{cognitive load} annotations. Moreover, AVCAffe is the largest affective computing dataset in English language. Additionally, we perform extensive analyses utilizing the self-reported ground-truths, which reveal interesting insights on the cognitive and affective states of participants during the experiment sessions. Given the sparsity of the annotations and the challenging nature of estimating cognitive load and affect, we introduce an interesting challenge to the deep learning research community. Furthermore, we present extensive baselines to provide benchmarks for future works. We believe AVCAffe would be a useful and challenging dataset for the deep learning community.


%% file: AAAI/camera_ready/supp.tex
\setcounter{table}{0}
\setcounter{figure}{0}
\renewcommand{\thetable}{S\arabic{table}}
\renewcommand\thefigure{S\arabic{figure}}


The organization of the supplementary material is as follows:
\begin{itemize}[noitemsep,nolistsep]
    \item Appendix \ref{supsec:availability}: Availability;
    \item Appendix \ref{supsec:questionnaire}: Questionnaire;
    \item Appendix \ref{supsec:baseline}: Details of Baselines;
    \item Appendix \ref{suppsec:inter-label}: Inter-label relationships;
    \item Appendix \ref{suppsec:borader_impact} Broader Impact;
    \item Appendix \ref{supsec:frames}: Additional Representative Frames.
\end{itemize}


\section{Availability} \label{supsec:availability}
This dataset is made freely available for the research community, which can be accessed from the project website.
The initial release of the dataset includes:
\begin{itemize}[noitemsep,nolistsep]
    \item[$\bullet$] Short video segments (average duration of $6$ seconds).
    \item[$\bullet$] {Face-crops corresponding to short video segments.}
    \item[$\bullet$] Full-length videos of each participant per task (video length of $2.5$-$10$ mins.).
    \item[$\bullet$] Self-reported ground truths for affect and cognitive load.
    \item[$\bullet$] Outcome of the pre-study questionnaire.
    \item[$\bullet$] Dataloader codes for easy and efficient use, written in PyTorch.
\end{itemize}


\section{Questionnaire} \label{supsec:questionnaire}
The questionnaire used to collect the self-reported ground truths are presented in Table \ref{tab:cog_form}. Please note, these questions are directly obtained from the original paper where NASA-TLX~\cite{nasa_tlx} and SAM~\cite{sam} are introduced for measuring cognitive load and affect respectively.


\section{Details of Baselines} \label{supsec:baseline}
To create the baselines, we use the official Pytorch \cite{Pytorch} implementations for all the audio and visual backbones. The MLP head for the multi-modal fusion networks consist of $2$ fully connected layers of hidden dimension $4096$ followed by ReLU \cite{relu} activation and dropout \cite{dropout}. Next, to train the networks, we downsample the visual stream at $8$ frames per second. Following, the facial crops are extracted using FaceNet \cite{facenet} to effectively classify affect and cognitive load states from the visual streams. Moreover, we resize the frames to a spatial resolution of $112^2$ and feed $2$ seconds of visual input to the visual encoder with a final input size of $3\!\times\!16\times\!112^2$. Next, we downsample the audio stream at $16$ kHz and use $2$ seconds of audio segments to the audio encoder. We transform the audio segments to mel-spectrograms using $80$ mel filters, set the hop size to $10$ milliseconds, and use an FFT window length of $1024$. Thus, the final audio input dimension becomes $80\!\times\!200$. 

Following, we apply standard augmentations on both audio and visual streams during the training. In particular, we apply {Multi-scale Crop}, {Random Horizontal Flip}, and {Color Jitter} on the visual segments. We then simply apply {Volume Jitter} on the audio waveforms. We train the baseline models with an Adam \cite{adam} optimizer for $20$ epochs using a warm-up multi-step learning rate scheduler with a batch size of $64$. Moreover, to tackle overfitting, we apply weight decay, dropout, and early stopping. To provide a wide range of baselines using different backbones, we sweep a range of hyperparameters and report the performance of the best models. Specifically, we try with learning rates \{0.00001, 0.00003, 0.00007, 0.00005, 0.0001\}, learning rate decays \{0.1, 0.5, 0.7\}, learning rate milestones \{(5, 10), (5, 15)\}, dropouts \{0.0, 0.5\}, and weight decays \{0.0, 1e-4, 1e-5, 1e-6\}. The uni-modal variants are trained on a single NVIDIA RTX6000 $24$ GB GPU, whereas the multi-modal variants are trained using $2$ GPUs in parallel.

\begin{table}[tb]
\centering
\fontsize{9pt}{10pt}\selectfont

\begin{tabular}{p{8cm}}
    \toprule
    \textbf{Cognitive Load:} Rate your response on a scale of 0-21 \\ \midrule \midrule
     \textbf{Mental Demand}: How mentally demanding was the task?   \\
     \textbf{Physical Demand}: How physically demanding was the task? \\ 
     \textbf{Temporal Demand}: How rushed was the pace of the task?  \\
     \textbf{Performance}: How successful were you in accomplishing what you were asked to do?  \\
     \textbf{Effort}: How hard did you have to work to accomplish your level of performance? \\ 
     \textbf{Frustration}: How insecure, discouraged, irritated, stressed, or annoyed were you? \\ 
     \bottomrule
\end{tabular}

\begin{tabular}{lp{7cm}}
    \multicolumn{2}{p{8cm}}{
    \textbf{Affect:} Choose the image in each category that best describes your state. }\\ \midrule \midrule
    \rotatebox{90}{Arousal} 
    & \includegraphics[width=0.39\textwidth]{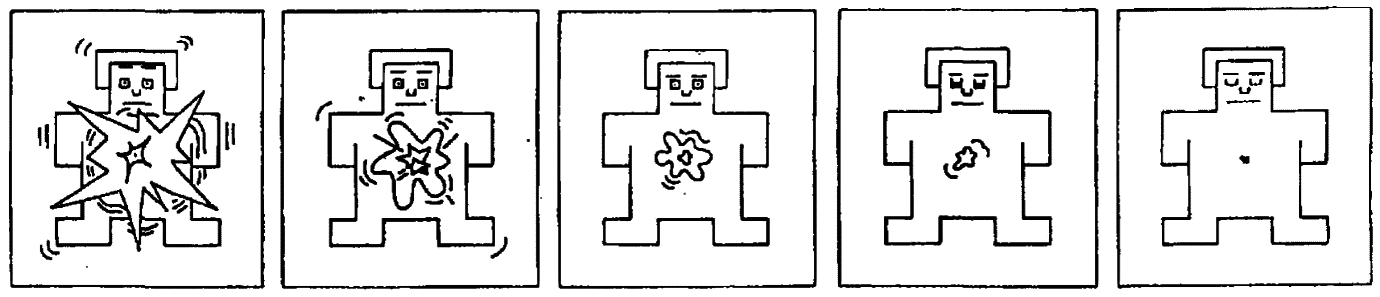} \\
    & {\small ~~Excited~Wide-awake~~Neutral~~~~~~~Dull~~~~~~~~~Calm}    \\ 
    \rotatebox{90}{Valence} 
    & \includegraphics[width=0.39\textwidth]{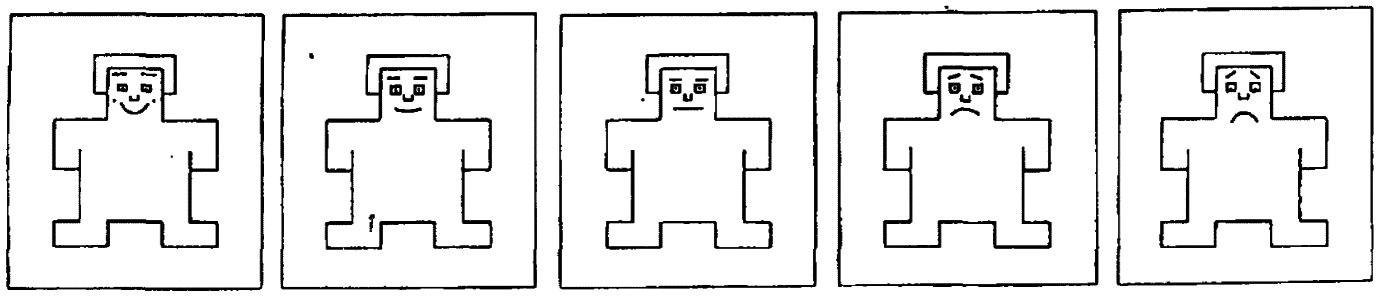} \\
    & {\small ~Pleasant~~~~Pleased~~~~~~Neutral~~Unsatisfied~~Unpleasant} 
\end{tabular}

\caption{{Cognitive load and affect questionnaires.} 
}
\label{tab:cog_form}
\end{table}

\section{Inter-label relationships} \label{suppsec:inter-label}
As discussed earlier, we explore the relationships between five output categories of affect and cognitive load attributes namely arousal, valence, mental demand, effort, and temporal demand. To provide a quantitative analysis between these output categories, we perform the Pearson Correlation test \cite{pearson} using the normalized self-reported scores. The results presented in Table \ref{tab:inter_label_corr} confirm our earlier qualitative findings, as it shows high correlations between effort and mental demand as well as effort and temporal demand, indicating that with increasing amounts of effort, participants experience higher mental and temporal load or vice-versa. 
Moreover, our statistical tests do not show any correlations between the cognitive load and affect attributes, which further confirms that our dataset has been able to successfully capture unique information beyond the common arousal and valence classes.

\begin{table}[t]
\fontsize{9pt}{10pt}\selectfont
\centering

\begin{tabular}{llr}
\toprule
Attribute 1 & Attribute 2 & Correlation \\ \midrule\midrule
Arousal  &  Valence  &  $ 0.321^* $ \\
Arousal  &  Effort  &  $ -0.095^* $ \\
Arousal  &  Mental Demand  &  $ -0.099^* $ \\
Arousal  &  Temporal Demand  &  $ -0.161^* $ \\
Valence  &  Effort  &  $ 0.179^* $ \\
Valence  &  Mental Demand  &  $ 0.201^* $ \\
Valence  &  Temporal Demand  &  $ 0.165^* $ \\
Effort  &  Mental Demand  &  $ 0.704^* $ \\
Effort  &  Temporal Demand  &  $ 0.485^* $ \\
Mental Demand  &  Temporal Demand  &  $ 0.463^* $ \\
\bottomrule
\end{tabular}%
\caption{Quantitative analysis on inter-label relationships. We present statistical correlations between different affect and cognitive load attributes. Statistical significance (denoted by $^*$) is considered at p $<0.01$.}
\label{tab:inter_label_corr}
\end{table}

\section{Broader Impact} \label{suppsec:borader_impact}
The proposed dataset would be of interest to researchers in both fields of psychology and computer science, to facilitate better understanding of cognitive load, affective states, and broadly human behaviors. The authors do not foresee any negative impacts. We also believe our work is very timely given the rise of remote work as one of the prominent paradigms of work in recent years. 


\section{Additional Representative Frames} \label{supsec:frames}
\begin{figure*}[!h]
    \centering
    \fontsize{9pt}{10pt}\selectfont
    \setlength\tabcolsep{0.2pt}
    \setlength{\arrayrulewidth}{4pt}
    \arrayrulecolor{tawny}
    \renewcommand{\arraystretch}{0.5}
    \resizebox{0.99\linewidth}{!}{
    \begin{tabular}{cccc}
    \includegraphics[width=0.45\textwidth]{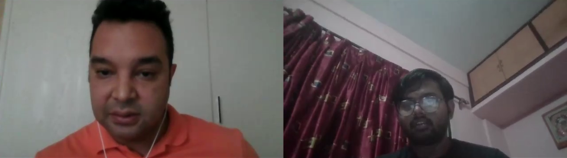} &
    \includegraphics[width=0.45\textwidth]{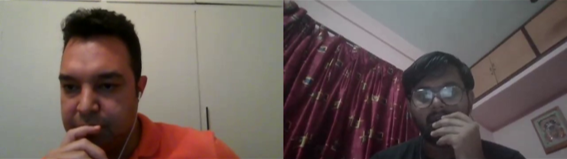} &
    \includegraphics[width=0.45\textwidth]{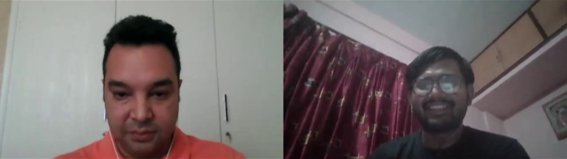} &
    \includegraphics[width=0.45\textwidth]{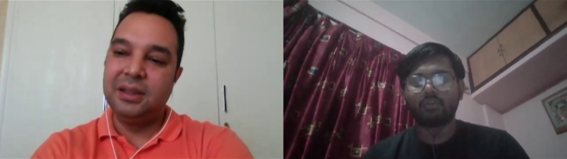} \\
    \includegraphics[width=0.45\textwidth]{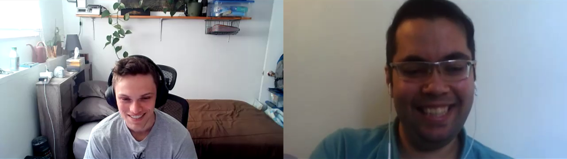} &
    \includegraphics[width=0.45\textwidth]{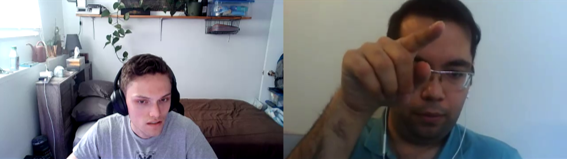} &
    \includegraphics[width=0.45\textwidth]{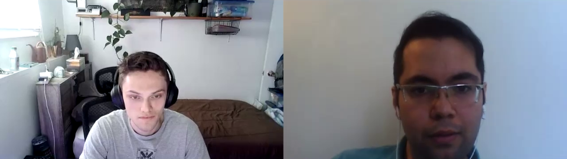} &
    \includegraphics[width=0.45\textwidth]{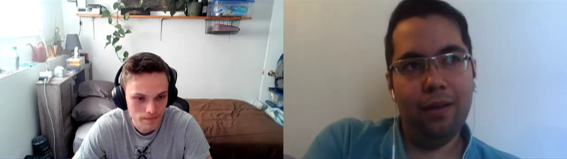} \\
    \includegraphics[width=0.45\textwidth]{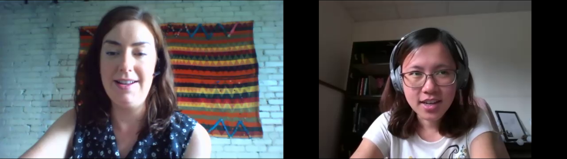} &
    \includegraphics[width=0.45\textwidth]{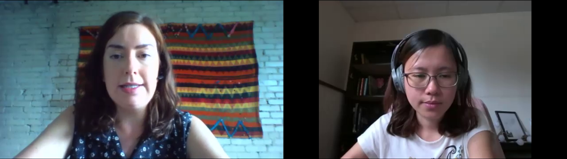} &
    \includegraphics[width=0.45\textwidth]{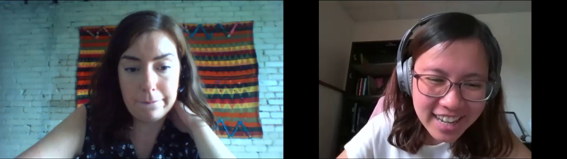} &
    \includegraphics[width=0.45\textwidth]{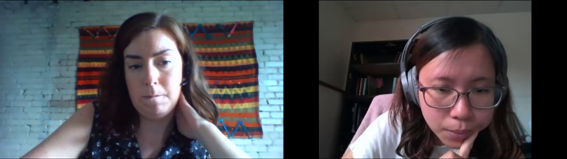} \\
    \includegraphics[width=0.45\textwidth]{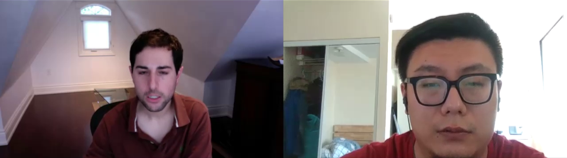} &
    \includegraphics[width=0.45\textwidth]{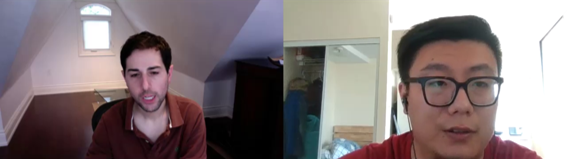} &
    \includegraphics[width=0.45\textwidth]{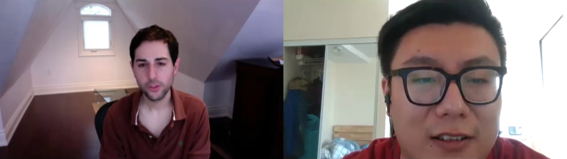} &
    \includegraphics[width=0.45\textwidth]{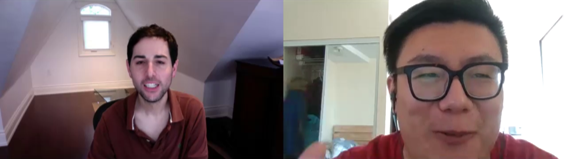} \\
    \end{tabular}
    }
    \caption{Sessions of participants from different ethnicity groups. Each row depicts a different session.
    }
    \label{fig:diff_ethnicity}
\end{figure*}

\begin{figure*}[!h]
    \centering
    \fontsize{9pt}{10pt}\selectfont
    \setlength\tabcolsep{0.2pt}
    \setlength{\arrayrulewidth}{4pt}
    \arrayrulecolor{tawny}
    \renewcommand{\arraystretch}{0.5}
    \resizebox{0.99\linewidth}{!}{
    \begin{tabular}{cccc}
    \includegraphics[width=0.45\textwidth]{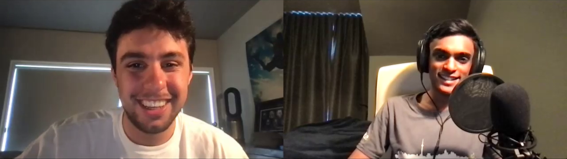} &
    \includegraphics[width=0.45\textwidth]{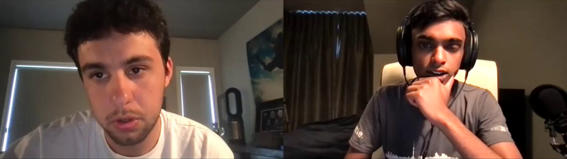} &
    \includegraphics[width=0.45\textwidth]{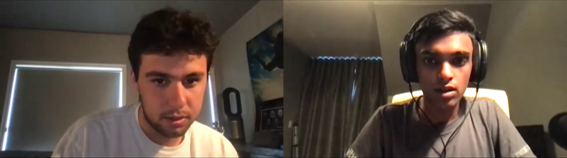} &
    \includegraphics[width=0.45\textwidth]{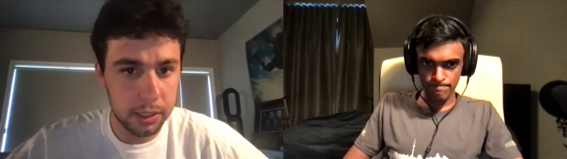} \\
    \includegraphics[width=0.45\textwidth]{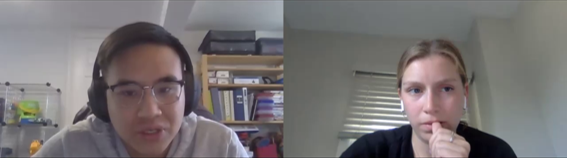} &
    \includegraphics[width=0.45\textwidth]{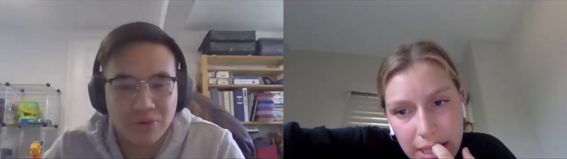} &
    \includegraphics[width=0.45\textwidth]{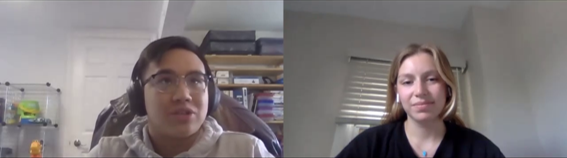} &
    \includegraphics[width=0.45\textwidth]{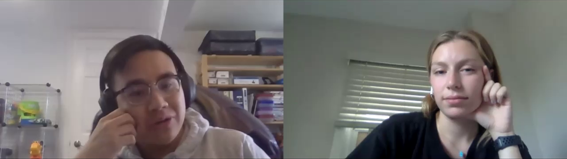} \\
    \includegraphics[width=0.45\textwidth]{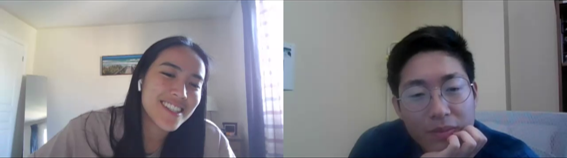} &
    \includegraphics[width=0.45\textwidth]{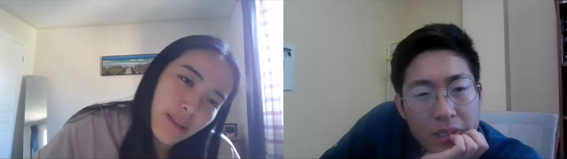} &
    \includegraphics[width=0.45\textwidth]{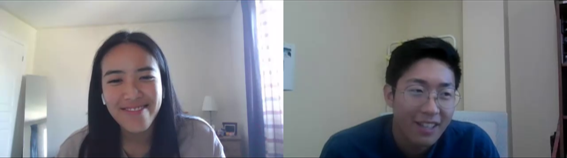} &
    \includegraphics[width=0.45\textwidth]{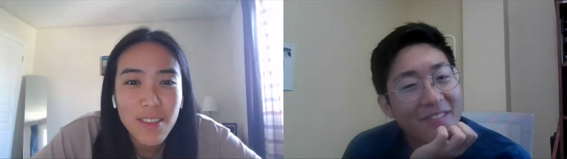} \\
    \includegraphics[width=0.45\textwidth]{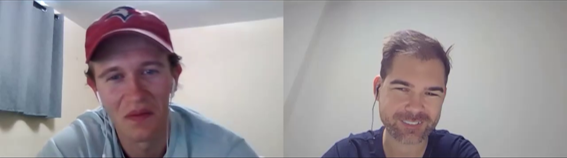} &
    \includegraphics[width=0.45\textwidth]{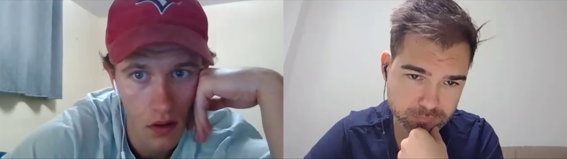} &
    \includegraphics[width=0.45\textwidth]{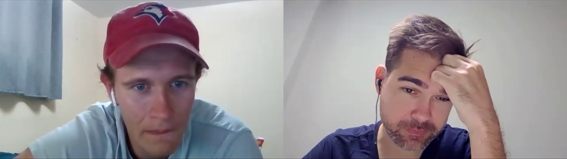} &
    \includegraphics[width=0.45\textwidth]{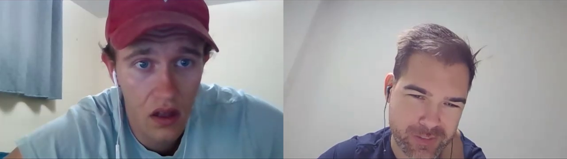} \\
    \end{tabular}
    }
    \caption{Sessions of participants from the same ethnicity groups. Each row depicts a different session.
    }
    \label{fig:same_ethnicity}
\end{figure*}

\begin{figure*}[!h]
    \centering
    \fontsize{9pt}{10pt}\selectfont
    \setlength\tabcolsep{0.2pt}
    \setlength{\arrayrulewidth}{4pt}
    \arrayrulecolor{tawny}
    \renewcommand{\arraystretch}{0.5}
    \resizebox{0.99\linewidth}{!}{
    \begin{tabular}{cccc}
    \includegraphics[width=0.45\textwidth]{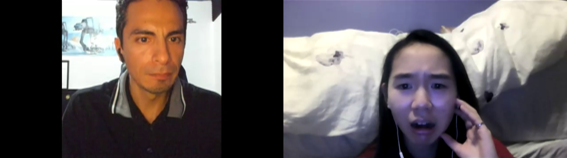} &
    \includegraphics[width=0.45\textwidth]{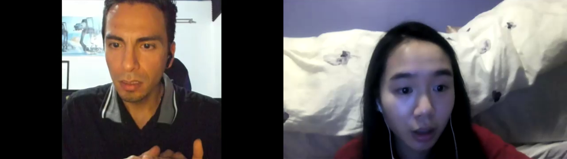} &
    \includegraphics[width=0.45\textwidth]{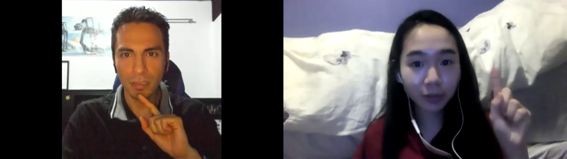} &
    \includegraphics[width=0.45\textwidth]{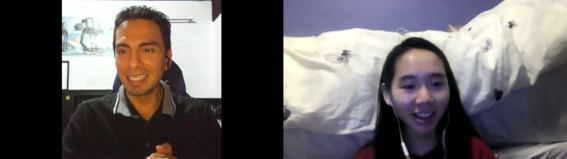} \\
    \includegraphics[width=0.45\textwidth]{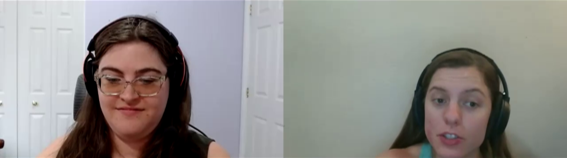} &
    \includegraphics[width=0.45\textwidth]{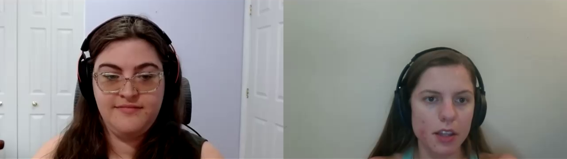} &
    \includegraphics[width=0.45\textwidth]{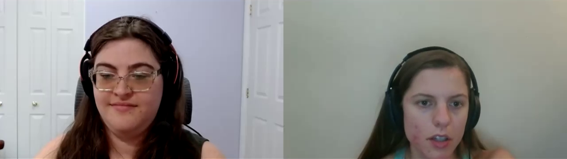} &
    \includegraphics[width=0.45\textwidth]{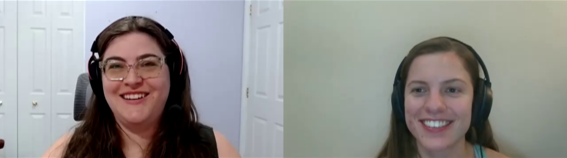} \\
    \includegraphics[width=0.45\textwidth]{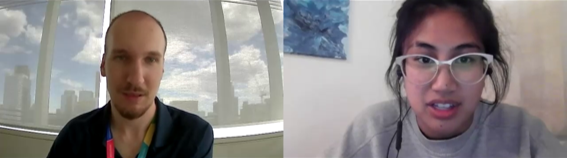} &
    \includegraphics[width=0.45\textwidth]{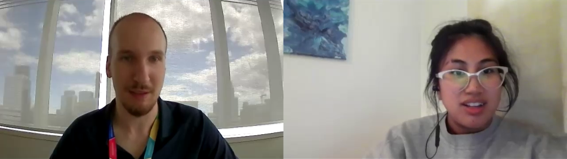} &
    \includegraphics[width=0.45\textwidth]{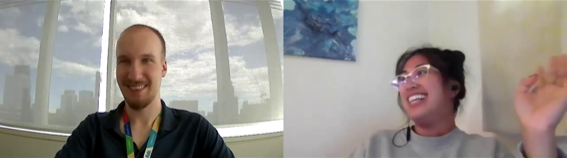} &
    \includegraphics[width=0.45\textwidth]{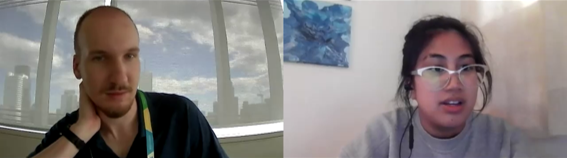} \\
    \includegraphics[width=0.45\textwidth]{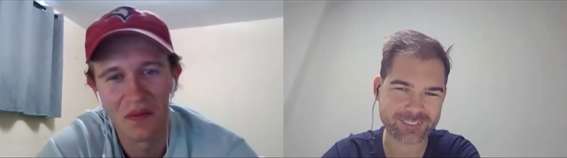} &
    \includegraphics[width=0.45\textwidth]{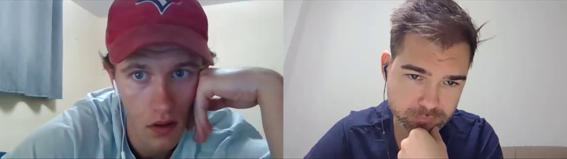} &
    \includegraphics[width=0.45\textwidth]{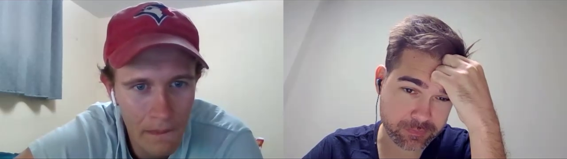} &
    \includegraphics[width=0.45\textwidth]{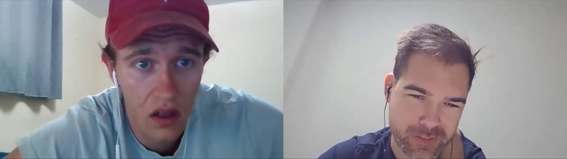} \\
    \end{tabular}
    }
    \caption{Sessions of participants from different age groups. Each row depicts a different session.
    }
    \label{fig:diff_age}
\end{figure*}

\begin{figure*}[!h]
    \centering
    \fontsize{9pt}{10pt}\selectfont
    \setlength\tabcolsep{0.2pt}
    \setlength{\arrayrulewidth}{4pt}
    \arrayrulecolor{tawny}
    \renewcommand{\arraystretch}{0.5}
    \resizebox{0.99\linewidth}{!}{
    \begin{tabular}{cccc}
    \includegraphics[width=0.45\textwidth]{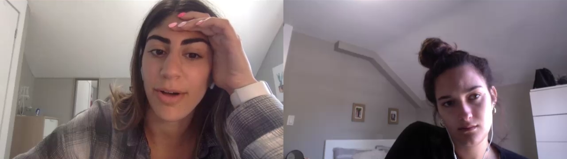} &
    \includegraphics[width=0.45\textwidth]{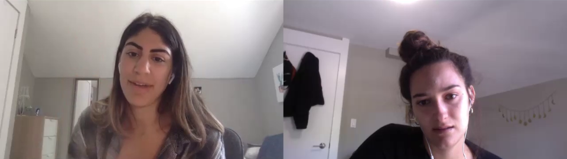} &
    \includegraphics[width=0.45\textwidth]{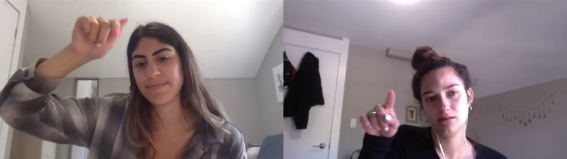} &
    \includegraphics[width=0.45\textwidth]{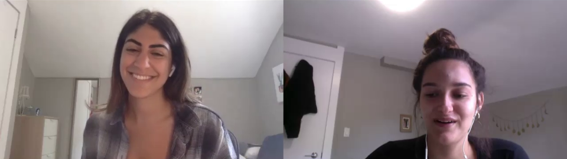} \\
    \includegraphics[width=0.45\textwidth]{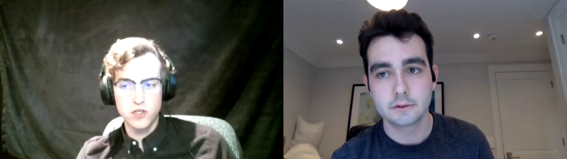} &
    \includegraphics[width=0.45\textwidth]{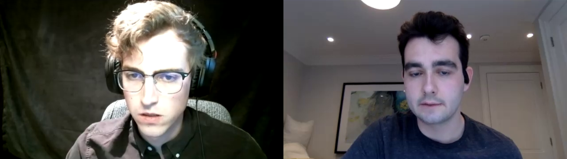} &
    \includegraphics[width=0.45\textwidth]{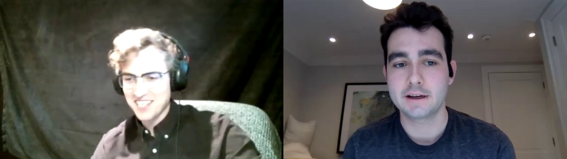} &
    \includegraphics[width=0.45\textwidth]{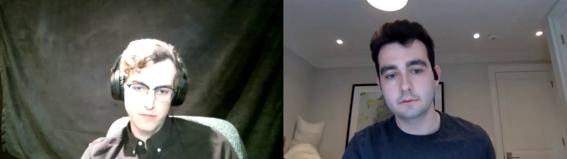} \\
    \includegraphics[width=0.45\textwidth]{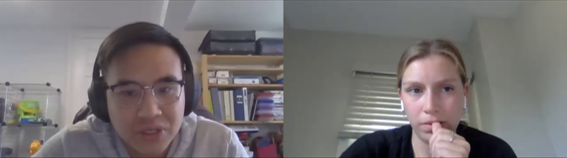} &
    \includegraphics[width=0.45\textwidth]{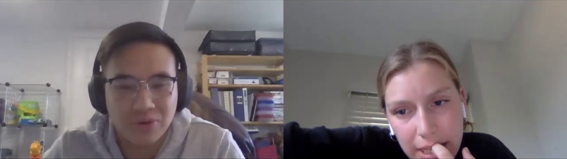} &
    \includegraphics[width=0.45\textwidth]{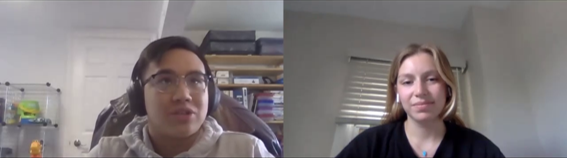} &
    \includegraphics[width=0.45\textwidth]{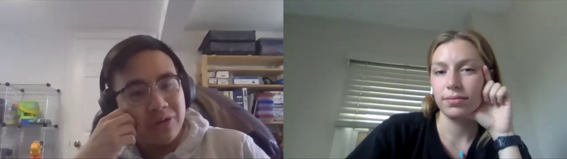} \\
    \includegraphics[width=0.45\textwidth]{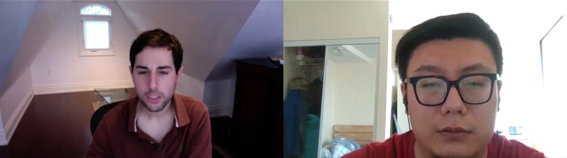} &
    \includegraphics[width=0.45\textwidth]{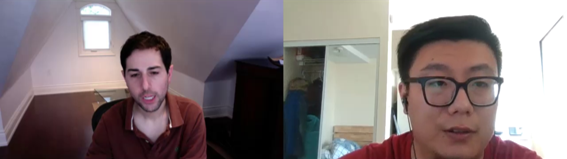} &
    \includegraphics[width=0.45\textwidth]{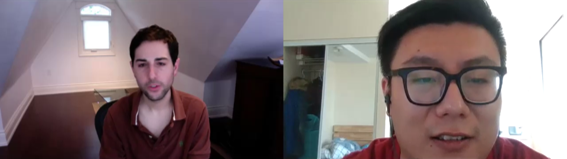} &
    \includegraphics[width=0.45\textwidth]{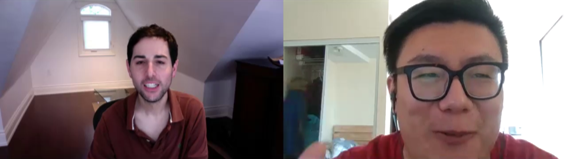} \\
    \end{tabular}
    }
    \caption{Sessions of participants from the same age groups. Each row depicts a different session.
    }
    \label{fig:same_age}
\end{figure*}

\begin{figure*}[!h]
    \centering
    \fontsize{9pt}{10pt}\selectfont
    \setlength\tabcolsep{0.2pt}
    \setlength{\arrayrulewidth}{4pt}
    \arrayrulecolor{tawny}
    \renewcommand{\arraystretch}{0.5}
    \resizebox{0.99\linewidth}{!}{
    \begin{tabular}{cccc}
    \includegraphics[width=0.45\textwidth]{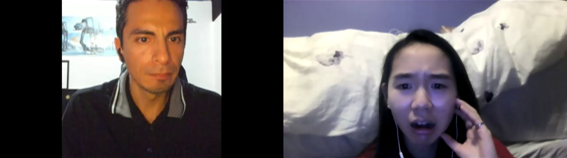} &
    \includegraphics[width=0.45\textwidth]{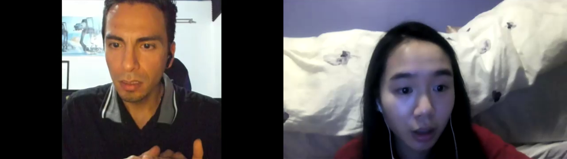} &
    \includegraphics[width=0.45\textwidth]{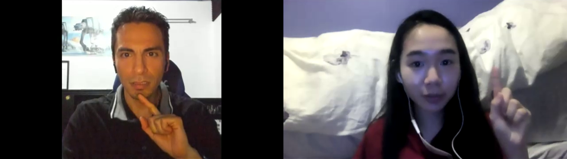} &
    \includegraphics[width=0.45\textwidth]{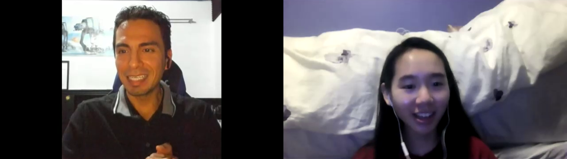} \\
    \includegraphics[width=0.45\textwidth]{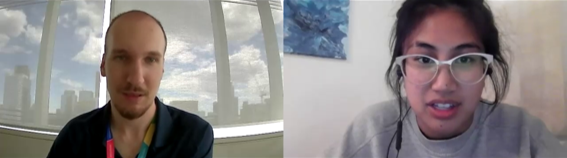} &
    \includegraphics[width=0.45\textwidth]{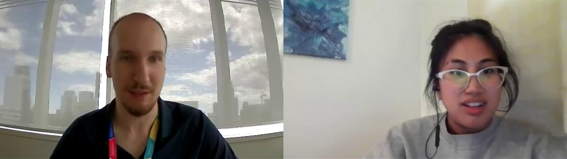} &
    \includegraphics[width=0.45\textwidth]{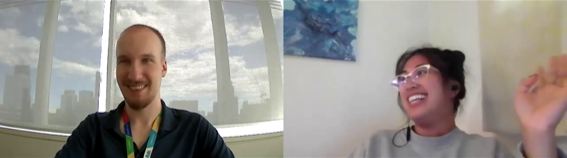} &
    \includegraphics[width=0.45\textwidth]{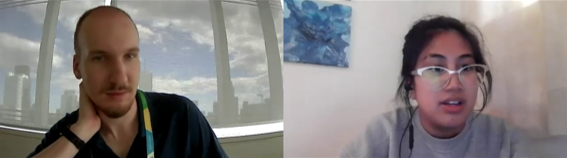} \\
    \includegraphics[width=0.45\textwidth]{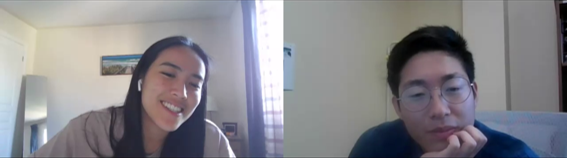} &
    \includegraphics[width=0.45\textwidth]{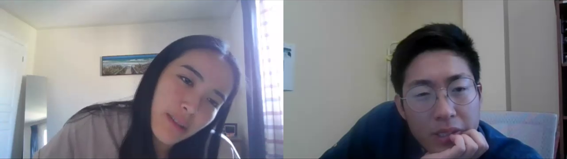} &
    \includegraphics[width=0.45\textwidth]{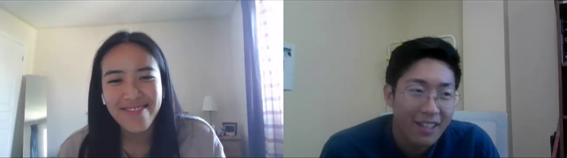} &
    \includegraphics[width=0.45\textwidth]{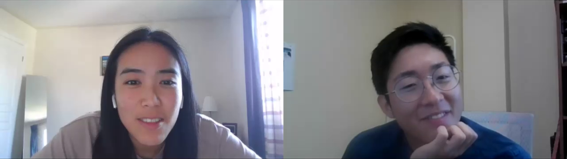} \\
    \includegraphics[width=0.45\textwidth]{figures/same_ethnicity/061_062_0.png} &
    \includegraphics[width=0.45\textwidth]{figures/same_ethnicity/061_062_1.png} &
    \includegraphics[width=0.45\textwidth]{figures/same_ethnicity/061_062_2.png} &
    \includegraphics[width=0.45\textwidth]{figures/same_ethnicity/061_062_3.png} \\
    \end{tabular}
    }
    \caption{Sessions of participants of different genders (female-to-male). Each row depicts a different session.
    }
    \label{fig:diff_gen}
\end{figure*}

\begin{figure*}[!h]
    \centering
    \fontsize{9pt}{10pt}\selectfont
    \setlength\tabcolsep{0.2pt}
    \setlength{\arrayrulewidth}{4pt}
    \arrayrulecolor{tawny}
    \renewcommand{\arraystretch}{0.5}
    \resizebox{0.99\linewidth}{!}{
    \begin{tabular}{cccc}
    \includegraphics[width=0.45\textwidth]{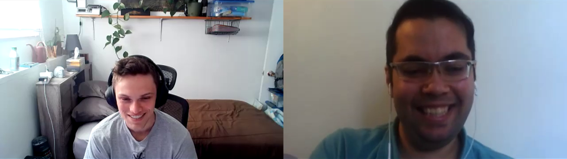} &
    \includegraphics[width=0.45\textwidth]{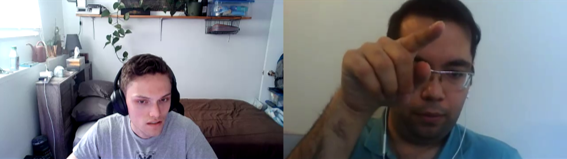} &
    \includegraphics[width=0.45\textwidth]{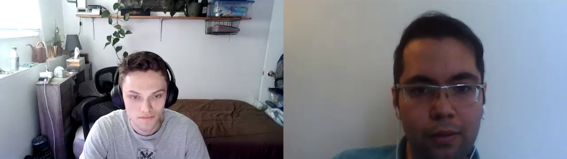} &
    \includegraphics[width=0.45\textwidth]{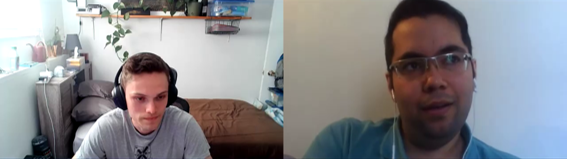} \\
    \includegraphics[width=0.45\textwidth]{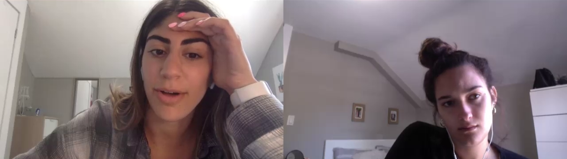} &
    \includegraphics[width=0.45\textwidth]{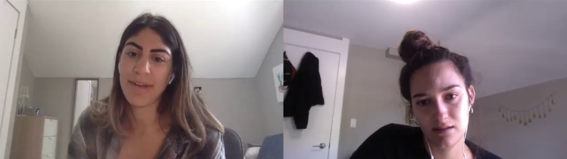} &
    \includegraphics[width=0.45\textwidth]{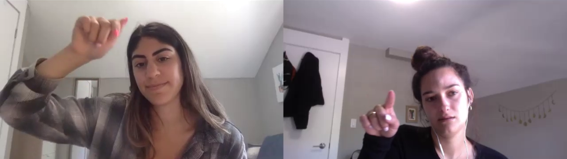} &
    \includegraphics[width=0.45\textwidth]{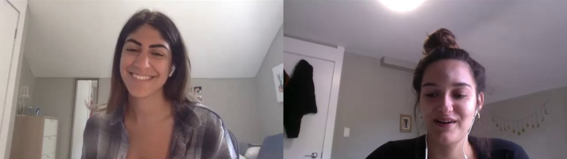} \\
    \includegraphics[width=0.45\textwidth]{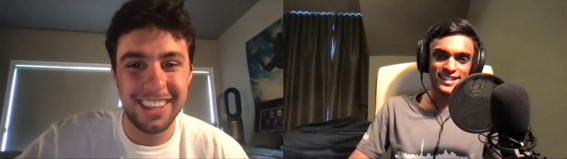} &
    \includegraphics[width=0.45\textwidth]{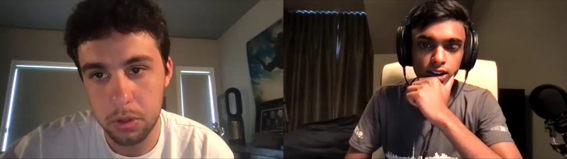} &
    \includegraphics[width=0.45\textwidth]{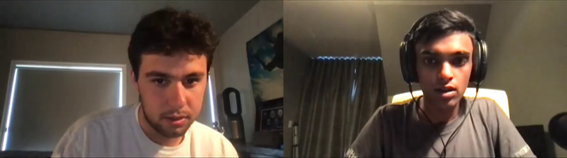} &
    \includegraphics[width=0.45\textwidth]{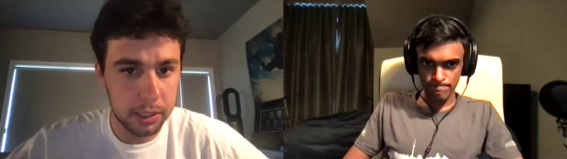} \\
    \includegraphics[width=0.45\textwidth]{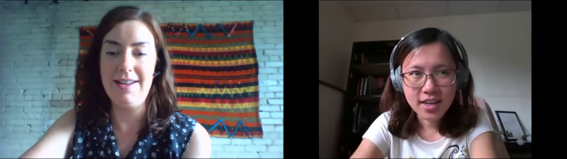} &
    \includegraphics[width=0.45\textwidth]{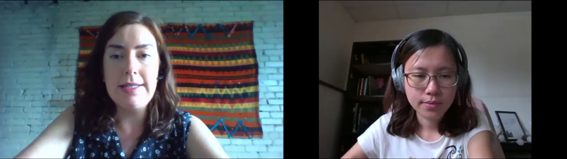} &
    \includegraphics[width=0.45\textwidth]{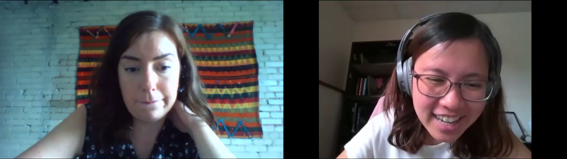} &
    \includegraphics[width=0.45\textwidth]{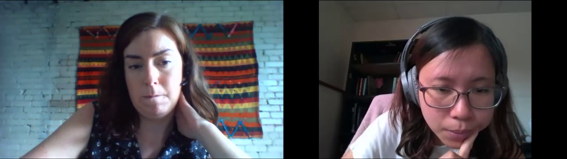} \\
    \end{tabular}
    }
    \caption{Sessions of participants of the same gender (male-to-male or female-to-female). Each row depicts a different session.
    }
    \label{fig:same_gen}
\end{figure*}

We present additional representative frames from different sessions in Figures \ref{fig:diff_ethnicity} through \ref{fig:same_gen}, showing the diversity of the participant pool in AVCAffe.

